# Quantum transport evidence of Weyl fermions in an epitaxial ferromagnetic oxide

Kosuke Takiguchi,[1,2,†] Yuki K. Wakabayashi,[1,*,†] Hiroshi Irie,[1] Yoshiharu Krockenberger,[1] Takuma Otsuka,[3] Hiroshi Sawada,[3] Sergey A. Nikolaev,[4,5] Hena Das,[4,5] Masaaki Tanaka,[2] Yoshitaka Taniyasu,[1] and Hideki Yamamoto[1]

[1]*NTT Basic Research Laboratories, NTT Corporation, Atsugi, Kanagawa 243-0198, Japan.*
[2]*Department of Electrical Engineering and Information Systems & Center for Spintronics Research Network (CSRN), The University of Tokyo, 7-3-1 Hongo, Bunkyo-ku, Tokyo 113-8656, Japan.*
[3]*NTT Communication Science Laboratories, NTT Corporation, Soraku-gun, Kyoto 619-0237, Japan.*
[4]*Laboratory for Materials and Structures, Tokyo Institute of Technology, 4259 Nagatsuta, Midori-ku, Yokohama, Kanagawa 226-8503, Japan.*
[5]*Tokyo Tech World Research Hub Initiative (WRHI), Institute of Innovative Research, Tokyo Institute of Technology, 4259 Nagatsuta, Midori-ku, Yokohama, Kanagawa 226-8503, Japan.*

*email: yuuki.wakabayashi.we@hco.ntt.co.jp
†These authors contributed equally to this work.

**Weyl fermions in a magnetic material have novel transport phenomena related to pairs of Weyl nodes[1,2,3,4,5], and they are of both scientific and technological interest, with the potential for use in high-performance electronics, spintronics and quantum computing. Although Weyl fermions have been predicted to exist in various oxides[6,7,8], evidence for their existence in oxide materials remains elusive[9,10,11]. $SrRuO_3$, a 4*d* ferromagnetic metal often used as an epitaxial conducting layer in oxide heterostructures[12,13,14,15], provides a promising opportunity to seek the existence of Weyl fermions in a magnetic material. State-of-the-art oxide thin film growth technologies, augmented by machine learning techniques, may allow access to such topological matter. Here we show direct quantum transport evidence of Weyl fermions in an epitaxial ferromagnetic oxide $SrRuO_3$: unsaturated linear positive magnetoresistance (MR)[16,17,18,19,20], chiral-anomaly-induced negative MR[1,16,21], $\pi$ Berry phase accumulated along cyclotron orbits[16,18,19,20], light cyclotron masses[16,17,18,19,20,22,23,24] and high quantum mobility of about 10000 $cm^2/Vs$[16,17,22,23,24,25,26,27]. We employed machine-learning-assisted molecular beam epitaxy (MBE)[28] to synthesize $SrRuO_3$ films whose quality is sufficiently high to probe their intrinsic quantum transport properties. We also clarified the disorder dependence of the transport of the Weyl fermions, and provided a brand-new diagram for the Weyl transport, which gives a clear guideline for accessing the topologically nontrivial transport phenomena. Our results establish $SrRuO_3$ as a magnetic Weyl semimetal and topological oxide electronics as a new research field.**

Weyl semimetals, which host Weyl fermions described by the Weyl Hamiltonian, have intriguing and fascinating transport phenomena based on the chiral anomaly and linear band dispersion with spin-momentum locking[1,2,3,4,29], such as chiral-anomaly-induced negative MR and high mobility[16,17,21]. Compared with space-inversion-symmetry-breaking Weyl semimetals[30], time-reversal-symmetry (TRS)-breaking ones are thought to be more suitable for spintronic applications[3,31,32]. For example, since the distribution of Weyl nodes in magnets is determined by the spin texture[1], this distribution is expected to be controlled by the magnetization switching technique[33,34]. Recent angle-resolved photoemission spectroscopy (ARPES) studies have found experimental evidence for the electronic structure of magnetic Weyl semimetals $Co_3Sn_2S_2$[2,3] and $Co_2MnGa$[4], such as the presence of bulk Weyl points with linear dispersions and surface Fermi arcs. Demonstrating the relevance of Weyl fermions in a magnetic material to spintronic and electronic applications requires information on quantum oscillations, which allows us to characterize transport properties of individual orbits in a magnetic Weyl semimetal. However, systematic and comprehensive measurements of quantum transport, including the quantum oscillations, have been hampered in magnetic Weyl semimetals because of the difficulty in achieving a quantum lifetime long enough to observe them in metallic systems. Since specimens in the form of epitaxial films are advantageous for future device applications of magnetic Weyl semimetals, it is urgently required to prepare single-crystalline thin films[35] of magnetic Weyl semimetals whose quality is sufficiently high to probe quantum transport properties.

Theoretically, the presence of Weyl fermions has been predicted for $SrRuO_3$, a 4*d* ferromagnetic material[7]. $SrRuO_3$ is widely used as an epitaxial conducting layer in oxide electronics and spintronics owing to the unique nature of ferromagnetic metal, compatibility with other perovskite-structured oxides, and chemical stability[12,13,14,15,34]. Theoretical studies predicted that the electronic structure of $SrRuO_3$ can includes a large number of Weyl nodes caused by the TRS breaking and spin-orbit coupling (SOC) (Fig. 1a)[7], and suggested that the Berry phase from the Weyl nodes gives rise to an anomalous Hall effect (AHE) in it.[7,10,36] However, a definitive conclusion on the presence of Weyl fermions near the Fermi energy ($E_F$) cannot be drawn from observations of the AHE alone[7,9,10], because the AHE reported so far for $SrRuO_3$ can be well reproduced using a function composed of both intrinsic (Karplus-Luttinger (KL) mechanism) and extrinsic (side-jump scattering) terms[37,38].

In this study, we conducted comprehensive high field magnetotransport measurements (see Methods section 'Magnetotransport measurements') including quantum oscillations of resistivity (i.e., Shubnikov-de Haas (SdH) oscillations), on an extraordinarily high-quality $SrRuO_3$ film (63-nm thick) epitaxially grown on $SrTiO_3$ (Fig. 1b, c and Extended Data Fig. 1a). Our first-principles electronic structure calculations predicted the presence of Weyl fermions within an energy range of −0.2 – 0.2 eV around the Fermi level in $SrRuO_3$. To probe the contribution of the Weyl fermions to the transport properties, it is necessary to identify the following five signatures from the magnetotransport data: (1) unsaturated linear positive MR[16,17,18,19,20] (2) chiral-anomaly-induced negative MR[1,16,21], (3) π Berry phase accumulated along the cyclotron orbits[16,18,19,20], (4) light cyclotron mass[16,18,19,20,22,23,24], and (5) high quantum mobility[16,17,22,23,24,25,26,27]. Although the light cyclotron mass, high mobility, and linear positive MR also exist in semiconductors with parabolic bands[39,40,41], we confirmed the existence of the Weyl fermions in $SrRuO_3$ by observing all the five signatures of Weyl fermions in $SrRuO_3$.

The residual resistivity ratio (RRR), which is defined as the ratio of the longitudinal resistivity $\rho_{xx}$ at 300 K [$\rho_{xx}$(300 K)] and $T$→0 K [$\rho_{xx}$($T$→0 K)] ($T$: temperature), is a good measure to gauge the purity of a metallic system, that is, the quality of single-crystalline $SrRuO_3$ thin films (See Methods section 'Determination of the RRR in $SrRuO_3$'). In fact, high RRR values are indispensable for exploring intrinsic electronic states. More specifically, RRR values above 40 and 60 have enabled observations of sharp, dispersive quasiparticle peaks near the Fermi level by ARPES[42] and quantum oscillations via electrical resistivity[43], respectively. To form $SrRuO_3$ with quality exceeding current levels, we employed our recently developed machine-learning-assisted MBE (see Methods section 'Machine-learning-assisted MBE')[28].

The resistivity $\rho_{xx}$ vs $T$ curve of the $SrRuO_3$ thin film shows a clear kink at the Curie temperature ($T_C$) of ~152 K (Fig. 1d)[12], while the magnetization measurement at $T$ = 10 K shows a typical ferromagnetic hysteresis loop (Fig. 1d, right inset). With a residual resistivity $\rho_{xx}(T\rightarrow 0\text{ K})$ of 2.23 μΩ·cm and an RRR of 84.3, $SrRuO_3$ thin films grown by machine-learning-assisted MBE are superior to those prepared by any other method[12,28,43,44]. Below approximately 20 K, the $T^2$ scattering rate ($\rho_{xx} \propto T^2$) expected in a Fermi liquid is observed (Fig. 1d, left inset)[12,42], indicating that the intrinsic transport phenomenon is seen below this temperature (hereafter called $T_F$).

In this Fermi liquid region [$T < T_F$ (20 K)], a semimetallic behaviour is seen in the MR ($\rho_{xx}(B)-\rho_{xx}(0\text{ T}))/\rho_{xx}(0\text{ T})$ (Fig. 1e) and Hall resistivity $\rho_{xy}(B)$ (Fig. 1f) curves with the magnetic field $B$ applied in the out-of-plane [001] direction of the $SrTiO_3$ substrate. As shown in Fig. 1e, $\rho_{xx}$ above $T_F$ shows the negative MR because of the suppression of magnetic scattering[12,45,46], and the MR changes its sign below $T_F$. Importantly, the positive MR at 2 K shows no signature of saturation even up to 14 T, which is typical of a semimetal[16,47] and also commonly seen in Weyl semimetals[1,16,17,18,19,20]. Especially in the case of Weyl semimetals, linear energy dispersion of Weyl nodes is considered to be one of the most plausible origins of unsaturated linear positive MR[48,49] (See Methods section "Excluding other possible origins of the positive MR in $SrRuO_3$"). In addition, as shown in Fig. 1f, the $\rho_{xy}(B)$ curves below $T_F$ are nonlinear, indicating the coexistence of multiple types of carriers (electrons and holes). We note that, below $T_F$, the AHE, which stems from the extrinsic side-jump scattering and intrinsic KL mechanisms in $SrRuO_3$,[37,38] is well suppressed due to the small residual resistivity of the $SrRuO_3$ film with the RRR of 84.3, and thus the $\rho_{xy}(B)$ curves below $T_F$ are dominated by the ordinary Hall effect (see Methods section "Temperature dependence of the AHE in the $SrRuO_3$ film with the RRR = 84.3"). Below 10 K, where the AHE is negligible, both the $\rho_{xy}(B)$ values and the slopes of $\rho_{xy}(B)$ change their signs in the high-$B$ region, signaling the possibility of the coexistence of high-mobility electrons with low-mobility holes[16]. Importantly, both the unsaturated linear positive MR and nonlinear Hall resistivity features start to appear simultaneously when the measurement temperature is decreased to the Fermi liquid range [$T < T_F$ (20 K)]. This indicates that the unsaturated linear positive MR stems from the electron- and hole-like Weyl fermions.

Next, we observed the chiral-anomaly-induced negative MR, which is an important signature of Weyl fermions[1,16,21,50]. To clarify the anisotropic character of the chiral-anomaly-induced negative MR thoroughly and systematically, we measured $\rho_{xx}(B)$ at $B$ angles $\alpha$, $\beta$, and $\gamma$ in the $xy$, $yz$, and $zx$ planes, respectively (Fig. 2a-c). The rotation angles $\alpha$, $\beta$, and $\gamma$ are defined in the insets of Fig. 2a-c. When $B$ is applied perpendicular to the current $I$ ($B \perp I$, $\alpha$ = 90° or $\beta$ = 0-90° or $\gamma$ = 90°), the unsaturated linear positive MR is observed. The unsaturated linear positive MR seen here is expected owing to the presence of the Weyl fermions, and those states are supposedly anisotropic because of the ~0.5% compressive strain of $SrRuO_3$ induced by the $SrTiO_3$ substrates[28]. This anisotropy is confirmed by varying $\beta$ (Fig. 2b). In contrast, when $B$ is rotated parallel to the current ($B//I$, $\alpha = 0°$ or $\gamma = 0°$), the MR turns negative and becomes linear above 8 T (Fig. 2a, c and Extended Data Fig. 4c). Theoretical calculations based on the semiclassical Boltzmann kinetic equation predict that TRS-breaking Weyl semimetals show a negative MR that is linear in $B$[1,50], in comparison with the quadratic dependence expected for space-inversion-symmetry-breaking Weyl semimetals[16,21]. Thus, the observed linear increase of the negative MR is consistent with a chiral anomaly in magnetic Weyl semimetals. The chiral-anomaly-induced negative MR can be understood from the violation of the conservation rules of chiral charges as shown in Fig. 2d[50], and it should be maximum when $B$ is exactly parallel to $I$ ($\alpha = 0°$, 180° or $\gamma = 0°$, 180°). As expected, this anisotropic feature of the negative MR is observed at $\alpha = 0°$, 180° or $\gamma = 0°$, 180° under 14 T (Fig. 2e). In addition, the peak structures in the angle dependence of the negative MR at $B//I$ ($\alpha = 0°$, 180° or $\gamma = 0°$, 180°) are similar to those in previous observations of the chiral anomaly in other Weyl semimetals[1,16,21,51,52]. These results confirm that this linear negative MR is induced by the chiral anomaly (See Methods section "Excluding other possible origins of the negative MR in $SrRuO_3$"). As a consequence of the contributions from the positive MR and negative MR, $\rho_{xx}(B)$ is lower than the zero field resistivity $\rho_0 = \rho_{xx}(0\ \mathrm{T})$ when $\alpha$ and $\gamma$ are near 0 or 180°.

The Berry phase has become an important concept in condensed matter physics over the past three decades, since it represents a topological classification of the system and it plays a fundamental role in various phenomena such as electric polarization, orbital magnetism, etc[53]. However, revealing the π Berry phase, originating from a band touching point of the Weyl node[16,], has been challenging in magnetic Weyl semimetals. Here, we detect the π Berry phase accumulation along the cyclotron orbits, for the first time in magnetic Weyl semimetals, by measurement techniques sensitive to the quantized energy levels, *i.e.*, SdH oscillation. The nontrivial π Berry phase, which is acquired by a surface integral of the Berry curvature Ω over a closed surface containing a Weyl point in the $k$-space (Fig. 3a)[16,18,19,20], causes a phase shift of quantum oscillations. According to the Lifshitz-Kosevich (LK) theory, the magnitude of the SdH oscillation is described as[16,18,19,20,54]

$$\Delta\sigma_{xx} = \sum_i A_i \frac{X_i}{\sinh X_i} \exp\left(-\frac{2\pi^2 k_B T_{Di}}{\hbar\omega_{ci}}\right) \cos\left[2\pi\left(\frac{F_i}{B} - \frac{1}{2} + \beta_{\mathrm{B}i}\right)\right] \qquad (1)$$

where $\Delta\sigma_{xx}$ is the oscillation component of the longitudinal conductivity, $A_i$ are the normalization factors, $X_i = 2\pi^2 k_{\mathrm{B}} T/\hbar\omega_{ci}$, $k_{\mathrm{B}}$ is the Boltzmann constant, $\hbar$ is the reduced Planck constant, $\omega_{ci}$ is the cyclotron frequency defined as $eB/m_i^*$, $m_i^*$ is the cyclotron mass, $T_{Di}$ is the Dingle temperature, $F_i$ is the frequency of the SdH oscillation, and $2\pi\beta_{\mathrm{B}i}$ is the phase shift caused by the Berry phase as mentioned above. The subscript $i$ is the label of an orbit of carriers. Note that, even when a large number of carriers in 3D multiband systems pin the chemical potential, the linear dispersion of the Weyl fermions leads to an unconventional $\pi$ phase shift in the SdH oscillation (See Methods section "Effect of the Fermi-level pinning on the phase shift in the SdH oscillation"). To extract the Berry phases in $SrRuO_3$, we used the LK formula [eq. (1)] for two frequencies to fit the SdH oscillation data. Figure 3b shows the SdH oscillation data at 2 K and the fitting results by eq. (1). (see Methods section 'Data pretreatment for quantum oscillations'). The oscillation spectrum is considerably complex because of the contribution from several subbands[7]. To reduce the fitting parameters, we first carried out fast Fourier transform of the SdH oscillation (Fig. 3c), and extracted the $F_1$ (26 T), $F_2$ (44 T), $m_1^*$ ($0.35m_0$), and $m_2^*$ (0.58$m_0$) ($m_0$, electron rest mass) values from the peak positions and the temperature dependence of $F_1$ and $F_2$ based on the LK theory (Fig. 3c, insets). Small masses are expected for Weyl fermions, which would fulfill the light cyclotron mass signature[16,18,19,20,22,23,24]. From the fitting to the data at 0.07 $T^{-1} < 1/B < 0.2$ $T^{-1}$ shown in Fig. 3b, we obtained $T_{D1}$ = 0.63 K, $T_{D2}$ = 0.34 K, $\beta_{B1}$ = 0.27, and $\beta_{B2}$ = 0.48. The $2\pi\beta_{B2}$ value of 0.96π indicates the presence of the nontrivial π Berry phase arising from the mass-less dispersion of the Weyl fermions (see Methods section "Exclusion of other possible mechanisms of the phase shift in $SrRuO_3$"). Although the interpretations of phase shifts between 0 and π in topological materials are still controversial, the $2\pi\beta_{B1}$ value of = 0.54π implies that the energy dispersion of the $F_1$ orbit has both quadratic (trivial) and linear (non-trivial) features[55,56]. These results cannot be reproduced by fixed zero Berry phases, confirming the existence of a non-zero Berry phase (Extended Data Fig. 6a).

The SdH oscillations not only give an insight into the topological nature, but also provide evidence of very high mobility of the Weyl fermions enclosing the Weyl points. We quantitatively determine the mobility of the charge carriers of the $F_1$ and $F_2$ orbits by calculating the quantum mobility, $\mu_{qi} = e\hbar/(2\pi k_B m_i^* T_{Di})$. The obtained $\mu_{q1}$ and $\mu_{q2}$ values are 9.6 × $10^3$ and 1.1×$10^4$ $cm^2/Vs$, respectively. In addition, assuming isotropic Weyl nodes, we can estimate the carrier concentrations $n_i = \frac{1}{6\pi^2}\left(\frac{2eF_i}{\hbar}\right)^{\frac{3}{2}}$ ($n_1$ = 3.8×$10^{17}$ $cm^{-3}$ and $n_2$ = 8.3×$10^{17}$ $cm^{-3}$) and the chemical potentials $\mu_i = e\hbar F_i/m_i^*$ ($\mu_1$ = 8.5 meV and $\mu_2$ = 8.8 meV).[57] These results mean that $F_1$ and $F_2$ come from the high-mobility and low-concentration Weyl fermions enclosing the Weyl points.

In addition to the Weyl fermions as evidenced by the five signatures described above, there are trivial Ru 4$d$ bands crossing the $E_F$ in $SrRuO_3$[7,12,42,43]. SdH experiments performed at 0.1 K (Fig. 3d) confirmed some of these trivial Fermi pockets with $F_3$ = 300 T, $F_4$ = 500 T, $F_5$ = 3500 T, and $F_6$ = 3850 T (see Methods section 'SdH oscillations of trivial orbits'). These four orbits have heavier masses (> 2.8 $m_0$) than those of $F_1$ and $F_2$, which are consistent with the reported values for $SrRuO_3$[12,43]. The Fermi pocket areas of $F_3$ (0.029 Å$^{-2}$) and $F_4$ (0.048 Å$^{-2}$) are close to those of the 364 T (0.035 Å$^{-2}$) orbit reported in earlier de Haas–van Alphen measurements[58], and the Fermi pocket areas of $F_5$ (0.334 Å$^{-2}$) and $F_6$ (0.368 Å$^{-2}$) correspond to the $\alpha_1$ band (0.33-0.37 Å$^{-2}$) observed by the ARPES and in earlier SdH measurements[42,43]. Noteworthy is that the Fermi pocket areas of $F_1$ (0.0025 Å$^{-2}$) and $F_2$ (0.0042 Å$^{-2}$) are more than ten times smaller than those of the trivial orbits, indicating that $F_1$ and $F_2$ stem from the small Fermi pockets as a feature of the Weyl fermions.

Observing the intrinsic transport signatures of the Weyl fermions requires a high-quality $SrRuO_3$ sample, since it is easily hindered by disorders such as defects and impurities. To show the disorder dependence of the transport phenomena in $SrRuO_3$, we investigated the RRR dependence of $T_C$, $T_F$, and the highest temperature where the linear positive MR, one of the clear features of Weyl fermion transport in $SrRuO_3$, remains (hereafter called $T_W$) (Fig. 4a-c) (see Methods section 'RRR dependence of the ferromagnetism, Fermi liquid behaviour, and Weyl behaviour'). As shown in Fig. 4c, the ferromagnetism becomes weaker ($T_C$ < 150 K) below RRR = 8.93, the Fermi liquid behaviour remains regardless of the RRR values even with low RRR = 6.61, and the positive MR is observed over the RRR = 19.4. It is remarkable that the positive MR ratio at 9 T increases with increasing RRR (Fig. 4a, b), which indicates that the Weyl fermions become more dominant in the transport. In the RRR dependence of the Hall resistivity, $\rho_{xy}(B)$, the nonlinear $B$ dependence becomes more prominent with a sign change in $d\rho_{xy}/dB$ with increasing RRR (see details in Methods section "RRR dependence of the Hall resistivity"). Thus, a high-quality $SrRuO_3$ sample is essential for observing the intrinsic transport of the Weyl fermions, and the diagram presented in Fig. 4c will be an effective guideline for realizing topologically non-trivial transport phenomena of the Weyl fermions to connect magnetic Weyl semimetals to spintronic devices[31,32].

Thus far, the magnetotransport data of a high-quality $SrRuO_3$ shows all of the expected marks of a magnetic Weyl semimetal. For certainty and theoretical rigor, we performed first-principles electronic structure calculations (see Methods section "Computational details") to analyze the energy dispersion of $SrRuO_3$. The calculated electronic structure for the orthorhombic *Pbnm* phase of $SrRuO_3$ for the ferromagnetic ground state shows a half-metallic behaviour (Fig. 5a) which agrees with previous electronic structure calculations[59]. The bands near the Fermi level are formed by the $t_{2g}$ Ru states hybridized with the O $2p$ orbitals (Extended Data Fig. 8a). The calculated magnetic moment per Ru ion (~ $1.4\mu_B$, as tabulated in Extended Data Table 2) is close to the experimental saturation magnetic moment of 1.25 $\mu_B$/Ru of our $SrRuO_3$ films.[28] We observe that in the ferromagnetic phase the Ru spins tend to align along the crystallographic $b$ axis reducing the symmetry of the system from $D_{2h}$ to $C_{2h}$. To identify the existence of the Weyl points (the band crossing points that carry $\pm 1$ chiral charge) by evaluating the outward Berry flux enclosed in a small sphere, we examined each band crossing of two pairs of bands I and II, shown in Fig. 5b, near the Fermi level in the presence of SOC with the magnetization along the orthorhombic $c$ axis. The resulting Weyl points are shown and listed in Fig. 5c and Extended Data Tables 3 and 4, respectively. We identified a total of 29 pairs of Weyl points with opposite chirality in the full Brillouin zone (BZ). An earlier theoretical study also made similar predictions of the existence of Weyl points in the case of a cubic structure of $SrRuO_3$.[7] Most of these Weyl points are found to exist within an energy range of −0.2 – 0.2 eV around the Fermi level. Among them, $|E - E_F|$ for $WP_{z}6_{1\text{-}4}$ (8–16 meV) in Extended Data Table 4 is very close to the experimental chemical potentials of the Weyl fermions estimated from the SdH oscillations ($\mu_1$ = 8.5 meV and $\mu_2$ = 8.8 meV for the $F_1$ and $F_2$ orbitals, respectively). The Weyl points near the Fermi level are expected to contribute to all the observed quantum transport phenomena reported in this study. It is important to note that a small monoclinic distortion induced by the substrate does not significantly change the band structure of $SrRuO_3$ and that Weyl points are robust with the distortion as long as the inversion symmetry is present (see Method section "First-principles calculations of Weyl points"), leading to a congruence of the experimental findings with theoretical predictions.

In conclusion, we have observed the emergence of Weyl fermions in epitaxial $SrRuO_3$ film with the best crystal quality ever reported[12], which was grown by machine-learning-assisted MBE. Experimental observation of the five important transport signatures of Weyl fermions—the linear positive MR, chiral-anomaly-induced negative MR, π phase shift in a quantum oscillation, light cyclotron mass, and high quantum mobility of about 10000 $cm^2/Vs$—combined with first-principles electronic structure calculations establishes $SrRuO_3$ as a magnetic Weyl semimetal. In addition, the RRR dependences of ferromagnetism, Fermi liquid behaviour, and positive MR serve as a road map to merge two emerging fields: topology in condensed matter and oxide electronics. Our results will pave the way for topological oxide electronics.

**Author contributions**
Y.K.W. conceived the idea, designed the experiments, and led the project. Y.K.W. and Y.K. designed and built the MBE system. Y.K.W., T.O., and H.S. implemented the Bayesian optimization algorithm for the sample growth. Y.K.W. grew the samples. Y.K.W. and K.T. carried out the sample characterizations. K.T., H.I., and Y.K.W. fabricated the Hall bar structures and carried out transport measurements. K.T. and Y.K.W. analysed and interpreted the data. S.A.N. and H.D. carried out the electronic-structure calculations. K.T., Y.K.W., H.I., Y.K., T.O., H.S., S.A.N., H.D., M.T., Y.T., and H.Y. contributed to the discussion of the data. K.T. and Y.K.W. co-wrote the paper with input from all authors.

**Data availability**
The data that support the plots in this paper and other finding of this study are available from the corresponding author upon request.

**Acknowledgements** We thank Yuichi Onuma for valuable discussions. A part of this work was conducted at Advanced Characterization Nanotechnology Platform of the University of Tokyo, supported by "Nanotechnology Platform" of the Ministry of Education, Culture, Sports, Science and Technology (MEXT), Japan.

**Author information** Correspondence and requests for materials should be addressed to Y.K.W. (yuuki.wakabayashi.we@hco.ntt.co.jp).

**References**

1. Kuroda, K. *et al.* Evidence for magnetic Weyl fermions in a correlated metal. *Nat. Mater.* **16**, 1090 (2017).
2. Morali, N. *et al.* Fermi-arc diversity on surface terminations of the magnetic Weyl semimetal $Co_3Sn_2S_2$. *Science* **365**, 1286 (2019).
3. Liu, D. F. *et al.* Magnetic Weyl semimetal phase in a Kagomé crystal. *Science* **365**, 1282 (2019).
4. Belopolski, I. *et al.* Discovery of topological Weyl fermion lines and drumhead surface states in a room temperature magnet. *Science* **365**, 1278 (2019).
5. Soh, J. R. *et al.* Ideal Weyl semimetal induced by magnetic exchange. *Phys. Rev. B* **100**, 201102(R) (2019).
6. Wan, X., Turner, A. M., Vishwanath, A. & Savrasov, S. Y. Topological semimetal and Fermi-arc surface states in the electronic structure of pyrochlore iridates. *Phys. Rev. B* **83**, 205101 (2011).
7. Chen, Y., Bergman, D. L. & Burkov, A. A. Weyl fermions and the anomalous Hall effect in metallic ferromagnets. *Phys. Rev. B* **88**, 125110 (2013).
8. Watanabe, H., Po, H. C. & Vishwanath, A. Structure and topology of band structures in the 1651 magnetic space groups. *Sci. Adv.* **4**, eaat8685 (2018).
9. Shimano, H. *et al*. Terahertz Faraday rotation induced by an anomalous Hall effect in the itinerant ferromagnet $SrRuO_3$. *Euro. Phys. Lett.* **95**, 17002 (2011).
10. Itoh, S. *et al.* Weyl fermions and spin dynamics of metallic ferromagnet $SrRuO_3$. *Nat. Commun.* **7**, 11788 (2016).

11. Ohtsuki, T. *et al.* Strain-induced spontaneous Hall effect in an epitaxial thin film of a Luttinger semimetal. *Proc. Natl. Acad. Sci.* **116**, 8803 (2019).
12. Koster, G. *et al.* Structure, physical properties, and applications of $SrRuO_3$ thin films. *Rev. Mod. Phys.* **84**, 253, (2012).
13. Worledge, D. C. & Geballe, T. H. Negative Spin-Polarization of $SrRuO_3$. *Phys. Rev. Lett.* **85**, 5182 (2000).
14. Boschker, H. *et al.* Ferromagnetism and Conductivity in Atomically Thin $SrRuO_3$. *Phys. Rev. X* **9**, 011027 (2019).
15. Matsuno, J. *et al.* Interface-driven topological Hall effect in $SrRuO_3$-$SrIrO_3$ bilayer. *Sci. Adv.* **2**, e1600304 (2016).
16 . Huang, X. *et al.* Observation of the chiral-anomaly-induced negative magnetoresistance: In 3D Weyl semimetal TaAs. *Phys. Rev. X* **5**, 031023 (2015).
17 . Singha, R. Pariari, A. K. Satpati, B. & Mandal, P. Large nonsaturating magnetoresistance and signature of nondegenerate Dirac nodes in ZrSiS. *Proc. Natl. Acad. Sci.* **114**, 2468 (2017).
18. He, L. P. *et al.* Quantum transport evidence for the three-dimensional dirac semimetal phase in $Cd_3As_2$. *Phys. Rev. Lett.* **113**, 246402 (2014).
19 . Xiang, Z. J. *et al.* Angular-Dependent Phase Factor of Shubnikov-de Haas Oscillations in the Dirac Semimetal $Cd_3As_2$. *Phys. Rev. Lett.* **115**, 226401 (2015).
20. Hu, J. *et al.* π Berry phase and Zeeman splitting of Weyl semimetal TaP. *Sci. Rep.* **6**, 18674 (2016).
21. Xiong, J. *et al*. Evidence for the chiral anomaly in the Dirac semimetal $Na_3Bi$. *Science* **350**, 413 (2015).
22. Zhang, C.L. *et al.* Magnetic-tunnelling-induced Weyl node annihilation in TaP. *Nat. Phys.* **13**, 979 (2017).
23. Arnold, F. *et al.*, Chiral Weyl Pockets and Fermi Surface Topology of the Weyl Semimetal TaAs. *Phys. Rev. Lett.* **117**, 146401 (2016).
24. Sergelius, P. *et al*. Berry phase and band structure analysis of the Weyl semimetal NbP. *Sci. Rep.* **6**, 33859 (2016).
25. Li, P. *et al*. Evidence for topological type-II Weyl semimetal $WTe_2$. *Nat. Commun.* **8**, 2150 (2017).
26 . Orbanić, F. *et al.* Three-dimensional Dirac semimetal and magnetic quantum oscillations in $Cd_3As_2$. *J. Phys. Conf. Ser.* **903**, 012038 (2017).
27. Takahashi, K. S. *et al.* Anomalous Hall effect derived from multiple Weyl nodes in high-mobility $EuTiO_3$ films. *Sci. Adv.* **4**, eaar7880 (2018).
28. Wakabayashi, Y. K. *et al.* Machine-learning-assisted thin-film growth: Bayesian optimization in molecular beam epitaxy of $SrRuO_3$ thin films. *APL Mater.* **7**, 101114 (2019).
29 . Fujioka, J. *et al.* Strong-correlation induced high-mobility electrons in Dirac semimetal of perovskite oxide. *Nat. Commun.* **10**, 362 (2019).
30. Xu, S. Y. *et al*. Discovery of a Weyl fermion semimetal and topological Fermi arcs. *Science* **349**, 613 (2015).
31. Kurebayashi, D. & Nomura, K. Voltage-Driven Magnetization Switching and Spin Pumping in Weyl Semimetals. *Phys. Rev. Appl.* **6**, 044013 (2016).

32. Araki, Y. & Nomura, K. Charge Pumping Induced by Magnetic Texture Dynamics in Weyl Semimetals. *Phys. Rev. Appl.* **10**, 014007 (2018).
33. Shiota, Y. *et al*. Induction of coherent magnetization switching in a few atomic layers of FeCo using voltage pulses. *Nat. Mater.* **11**, 39 (2012).
34. Liu, L. *et al*. Current-induced magnetization switching in all-oxide heterostructures. *Nat. Nanotechnol.* **14**, 939 (2019).
35. Ma, Y. *et al*. Realization of Epitaxial Thin Films of the Topological Crystalline Insulator $Sr_3SnO$. Preprint at https://arxiv.org/abs/1912.13431 (2019).
36. Fang, Z. *et al.* The Anomalous Hall Effect and Magnetic Monopoles in Momentum Space. *Science* **302**, 92 (2003).
37. Karplus, R. & Luttinger, J. Hall Effect in Ferromagnetics. *Phys. Rev.* **95**, 1154 (1954).
38. Haham, N. *et al.* Scaling of the anomalous Hall effect in $SrRuO_3$. *Phys. Rev. B* **84**, 174439 (2011).
39. Khouri, K. *et al*. Linear Magnetoresistance in a Quasifree Two-Dimensional Electron Gas in an Ultrahigh Mobility GaAs Quantum Well. *Phys. Rev. Lett.* **117**, 256601 (2016).
40. Sammon, M. Zudov, M. A., and Shklovskii, B. I. Mobility and quantum mobility of modern GaAs/AlGaAs heterostructures. *Phys. Rev. Mater.* **2**, 064604 (2018).
41. Madelung, O *Semiconductors; Data Handbook* (Springer, Berlin, Germany 2013)
42. Shai, D. E. *et al.* Quasiparticle Mass Enhancement and Temperature Dependence of the Electronic Structure of Ferromagnetic $SrRuO_3$ Thin Films. *Phys. Rev. Lett.* **110**, 087004 (2013).
43. Mackenzie, A. & Reiner, J. Observation of quantum oscillations in the electrical resistivity. *Phys. Rev. B.* **58**, R13318(R) (1998).
44. Izumi, M. *et al.* Magnetotransport of $SrRuO_3$ thin film on $SrTiO_3$ (001). *J. Phys. Soc. Jpn.* **66**, 3893 (1997).
45. Coey, L. M. D. & Venkatesan, M. Half-metallic ferromagnetism : Example of $CrO_2$. *J. Appl. Phys.* **91**, 8345 (2012).
46. Wang, K. Y., Edmonds, K. W., Campion, R. P., Zhao, L. X., Foxon, C. T. & Gallagher, B. L. Anisotropic magnetoresistance and magnetic anisotropy in high-quality (Ga,Mn)As films. *Phys. Rev. B* **72**, 085201 (2005).
47. Ali, M. N. et al. Large, non-saturating magnetoresistance in $WTe_2$. *Nature* **514**, 205 (2014).
48. Abrikosov, A., Quantum magnetoresistance. *Phys. Rev. B* **58**, 2788 (1998).
49. Shekhar, C. et al. Extremely large magnetoresistance and ultrahigh mobility in the topological Weyl semimetal candidate NbP. *Nat. Phys.* **11**, 645 (2015).
50. Son, D. T. & Spivak, B. Z. Chiral anomaly and classical negative magnetoresistance of Weyl metals. *Phys. Rev. B* **88**, 104412 (2013).
51. Zhang, C.-L. *et al*. Signatures of the Adler-Bell-Jackiw chiral anomaly in a Weyl fermion semimetal. *Nat. Commun*. **7**, 10735 (2016).
52. Hirschberger, M. *et al*. The chiral anomaly and thermopower of Weyl fermions in the half-Heusler GdPtBi. *Nat. Mater.* **15**, 1161 (2016).
53. Xiao, D., Chang, M. C. & Niu, Q. Berry phase effects on electronic properties. *Rev. Mod. Phys.* **82**, 1959 (2010).
54. Murakawa H. *et al.* Detection of Berry's Phase in a Bulk Rashba Semiconductor.

*Science* **342**, 1490 (2013).
55. Wright, A. R. & McKenzie, R. H. Quantum oscillations and Berry's phase in topological insulator surface states with broken particle-hole symmetry. *Phys. Rev. B* **87**, 085411 (2013).
56. LuK'Yanchuk, I. A. & Kopelevich, Y. Phase analysis of quantum oscillations in graphite. *Phys. Rev. Lett.* **93**, 166402 (2004).
57. Eto, K., Ren, Z., Taskin, A. A., Segawa, K. & Ando, Y. Angular-dependent oscillations of the magnetoresistance in $Bi_2Se_3$ due to the three-dimensional bulk Fermi surface. *Phys. Rev. B* **81**, 195309 (2005).
58. Alexander, C. S., McCall, S., Schlottmann, P., Crow, J. E. & Cao, G. Angle-resolved de Haas-van Alphen study of $SrRuO_3$. *Phys. Rev. B* **72**, 024415 (2005).
59. Ryee, S., Jang, S. W., Kino, H., Kotani, T. & Han, M. J. Quasiparticle self-consistent GW calculation of $Sr_2RuO_4$ and $SrRuO_3$. *Phys. Rev. B* **93**, 075125 (2016).

## Figures and figure legends

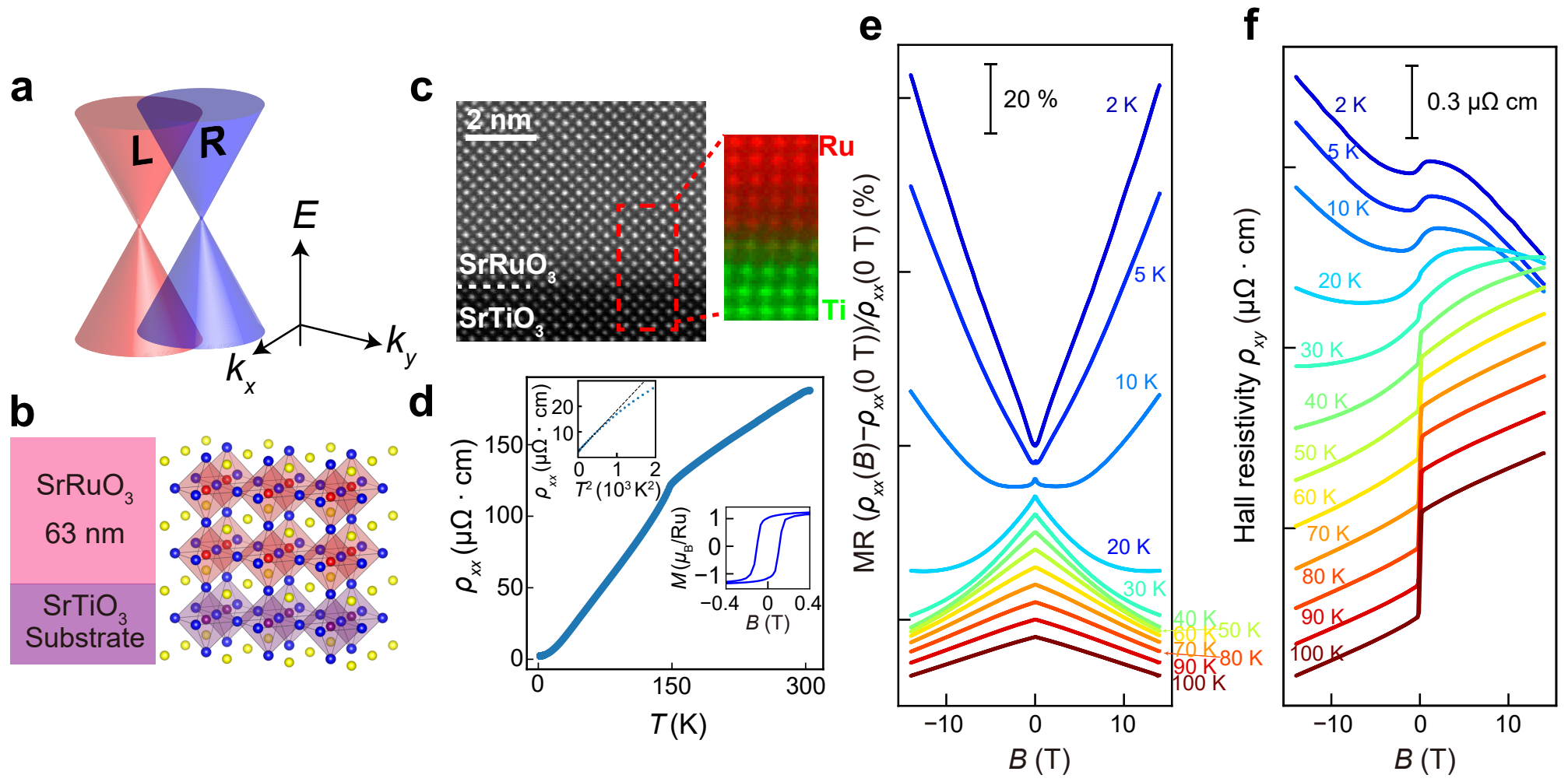

**Fig. 1 Sample characteristics and temperature dependence of the magnetoresistance and Hall resistivity of $SrRuO_3$. a,** Schematic image of a pair of Weyl nodes with opposite chiralities (L and R). In FM $SrRuO_3$, the TRS breaking lifts the spin degeneracy and leads to linear band crossing at many $k$ points, resulting in a pair of Weyl nodes with opposite chiralities. **b,** Schematic of the sample and crystal structures of the $SrRuO_3$ films (63-nm thick) on a $SrTiO_3$ substrate. In the schematic crystal image, yellow, blue, red, and purple spheres indicate strontium, oxygen, ruthenium, and titanium, respectively. **c,** Cross-sectional high-angle annular dark field scanning transmission electron microscopy (HAADF-STEM) image of the $SrRuO_3$ film with the RRR of 71 taken along the [100] axis of the $SrTiO_3$ substrate. The inset in **c** shows a color overlay of the electron energy loss spectroscopy (EELS)-STEM images for the Ti-$L_{2,3}$- (green) and Ru-$M_{4,5}$-edge (red). Epitaxial growth of the high-quality single-crystalline $SrRuO_3$ film with an abrupt substrate/film interface is seen in the images. **d,** $\rho_{xx}$-$T$ curves for the $SrRuO_3$ film with the RRR of 84.3. The left inset in **d** shows the $\rho_{xx}$ versus $T^2$ plot with the linear fitting (black dashed line). We defined the Fermi liquid region as the temperature range where the experimental $\rho_{xx}$ and the fitting line are close enough to each other (< 0.1 μΩ cm). The right inset in **d** shows the magnetization $M$ versus $B$ curve at 10 K with $B$ applied in the out-of-plane [001] direction of the $SrTiO_3$ substrate. **e, f,** MR ($\rho_{xx}(B)-\rho_{xx}(0\ T))/\rho_{xx}(0\ T)$ and Hall resistivity $\rho_{xy}(B)$ curves at 2 to 100 K for the $SrRuO_3$ film with the RRR of 84.3 with $B$ applied in the out-of-plane [001] direction of the $SrTiO_3$ substrate. In **e** and **f**, the MR and the Hall resistivity at each temperature have been offset by 7% and 0.22 μΩ·cm, respectively, for easy viewing.

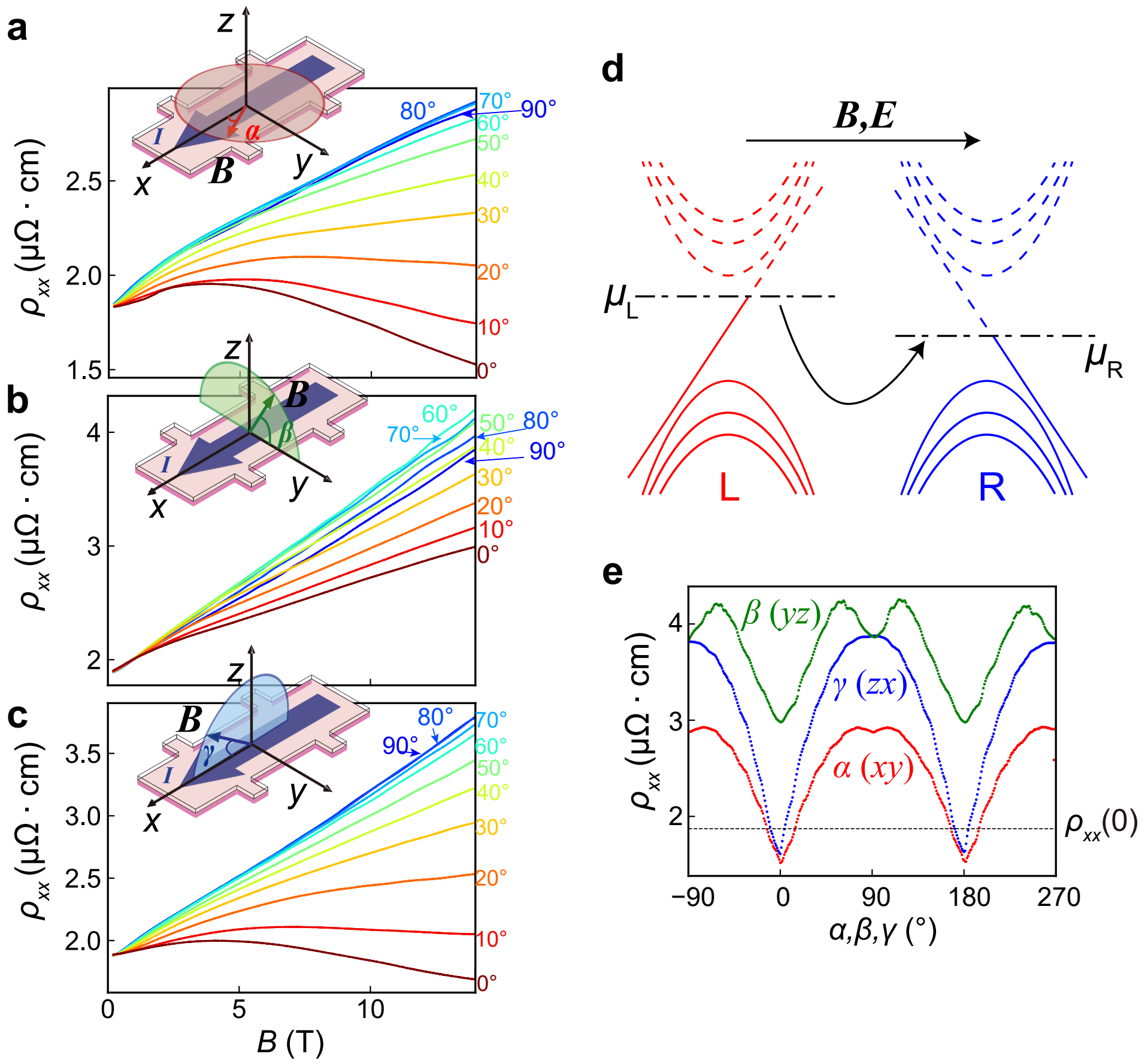

**Fig. 2 Chiral anomaly in the transport of $SrRuO_3$. a-c**, MR $\rho_{xx}(B)$ at $B$ angles $\alpha$ (**a**), $\beta$ (**b**), and $\gamma$ (**c**) at 2 K for the $SrRuO_3$ film with the RRR of 84.3. Angles $\alpha$, $\beta$, and $\gamma$ are defined in the insets of **a-c**. **d**, Schematic image of the chiral-anomaly-induced negative MR. Red and blue bands represent Landau levels of a pair of Weyl nodes with opposite chiralities L and R, respectively. Non-orthogonal electric and magnetic fields ($\boldsymbol{E}\cdot\boldsymbol{B} \neq 0$) lead to the chiral charge transfer between the two Weyl nodes with opposite chiralities (L and R). The currents of the chiral charges are observed as a form of negative MR. Here, $\mu_L$ and $\mu_R$ indicate chemical potentials in the Weyl points with opposite chiralities. **e**, $\alpha$-, $\beta$-, and $\gamma$-angular dependences of the MR $\rho_{xx}$ with $B$ = 14 T at 2 K for the $SrRuO_3$ film with the RRR of 84.3. The black dashed line indicates the $\rho_{xx}$(0 T) value. The region below the black dashed line shows negative MR.

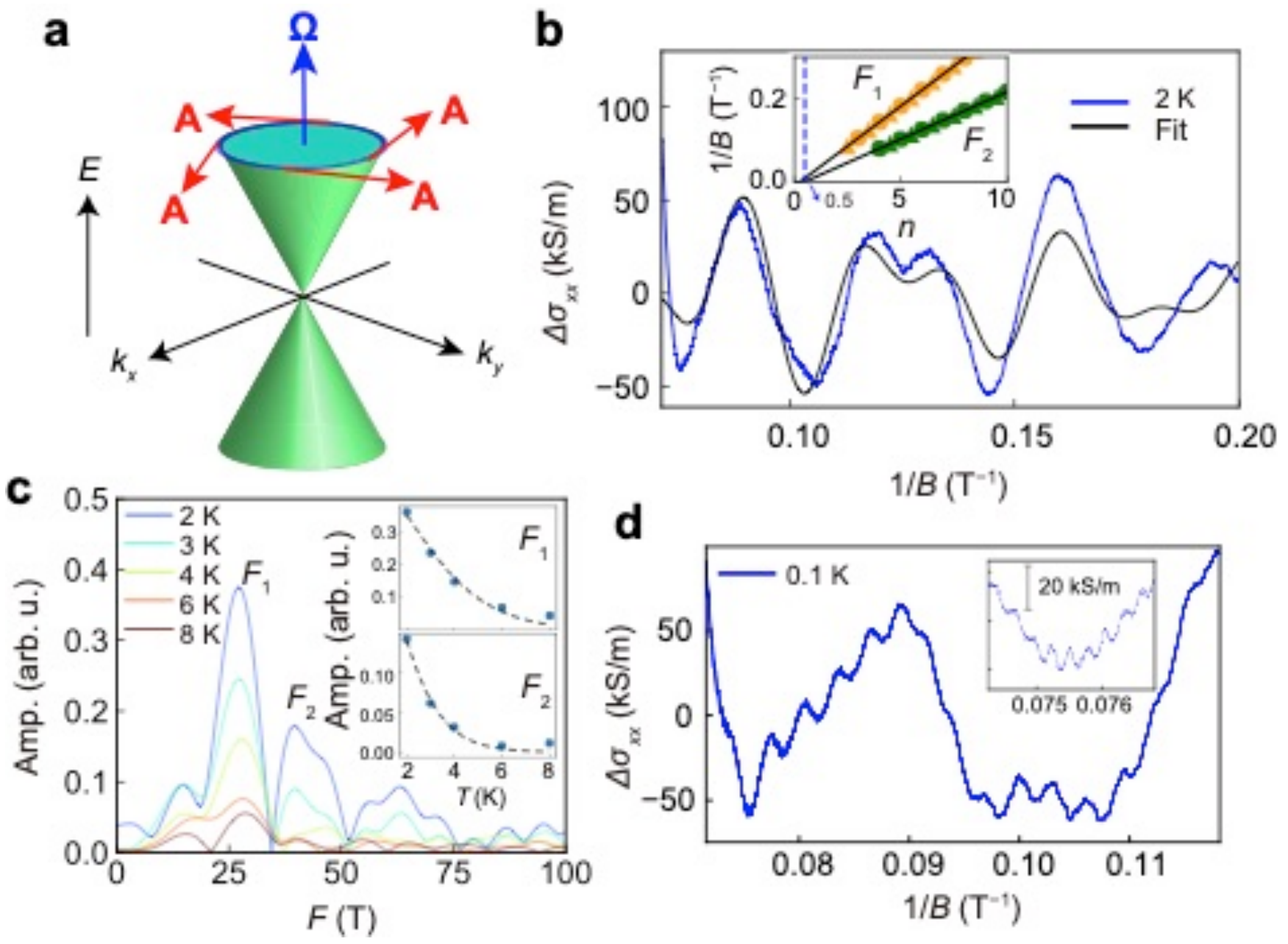

**Fig. 3 Quantum oscillations of $SrRuO_3$.** **a**, Schematic image of energy dispersion near a Weyl point. Blue and red arrows represent the Berry curvature Ω and the Berry connection A, respectively. The π Berry phase is accumulated along the cyclotron orbits (blue circle). **b**, SdH oscillation measured at 2K with $B$ (5 T < $B$ < 14 T) applied in the out-of-plane [001] direction of the $SrTiO_3$ substrate for the $SrRuO_3$ film with the RRR of 84.3. Black curve is the fitting curve of eq. (1). The fitting was carried out by a non-linear least squares method with the fitting parameters $A_1$, $\beta_1$, $T_{D1}$, $A_2$, $\beta_2$, and $T_{D2}$. The inset shows fan diagrams of two oscillation components of $F_1$ (= 26 T) and $F_2$ (= 44 T), which are shown as orange and green symbols, respectively. Here, the circles and triangles indicate integer and half-integer indexes of the oscillation components. **c**, Fourier transform spectra of the SdH oscillations at 2 to 8 K. Insets in **c** show the mass estimations of the $F_1$ and $F_2$ orbits according to the LK theory. Black dashed curves are the fitting curves. **d**, SdH oscillation observed at 0.1 K for the $SrRuO_3$ film with the RRR of 84.3. The inset in **d** shows a close-up at around 0.075 $T^{-1}$. The oscillation holds four kinds of other trivial orbits with higher frequencies.

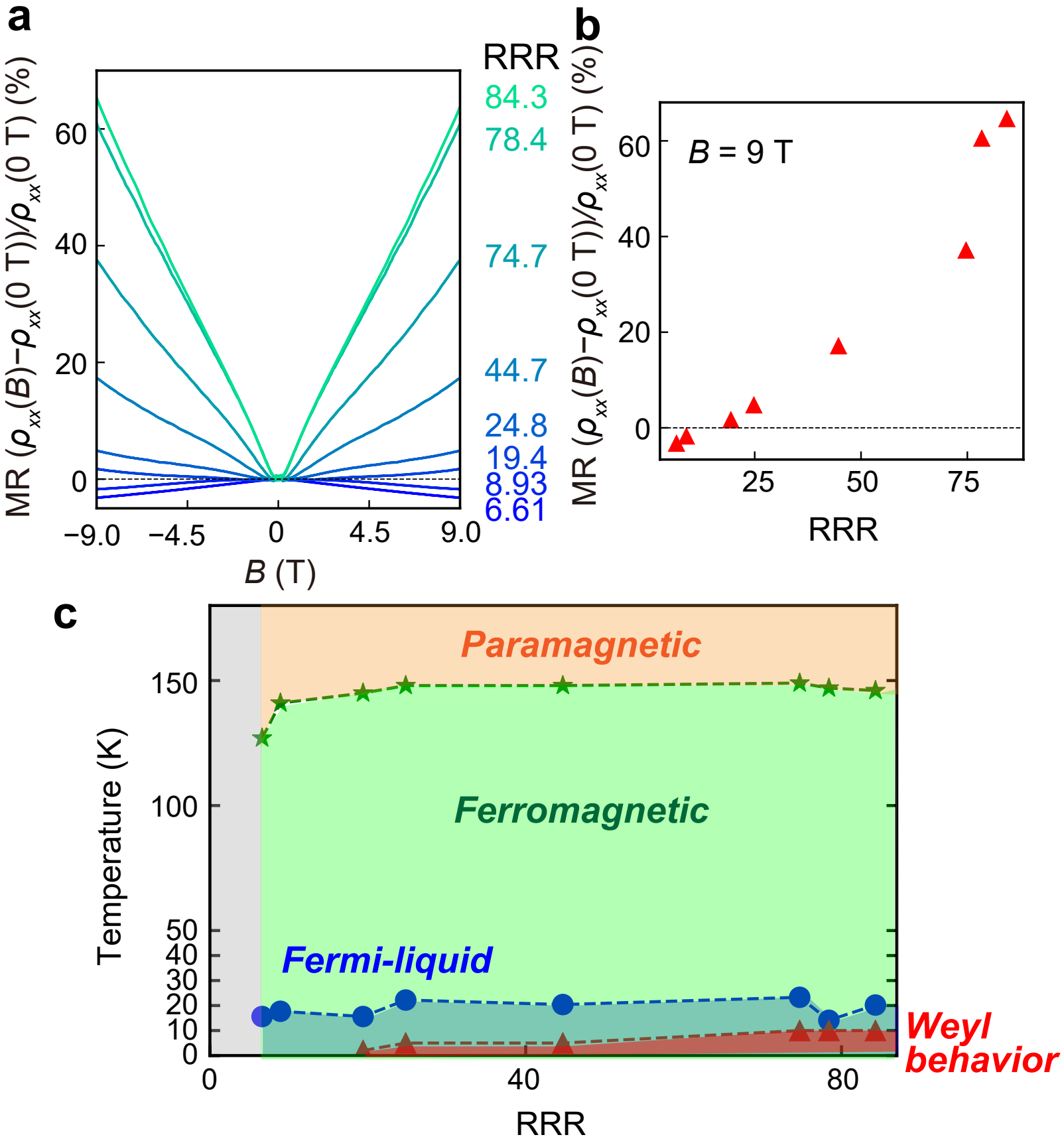

**Fig. 4 RRR dependence of the transport phenomena in $SrRuO_3$. a**, RRR dependence of the MR $(\rho_{xx}(B)-\rho_{xx}(0\ \mathrm{T}))/\rho_{xx}(0\ \mathrm{T})$ measured at 2 or 2.3 K. When RRR > 19.4, the positive MR, which is one piece of experimental evidence of the Weyl fermions in $SrRuO_3$, clearly emerges. **b**, RRR dependence of the MR $(\rho_{xx}(B)-\rho_{xx}(0\ \mathrm{T}))/\rho_{xx}(0\ \mathrm{T})$ at $B$ = 9 T and $T$ = 2 or 2.3 K. **c**, Diagram of the RRR dependence of $T_C$, $T_F$, and $T_W$, which are shown as green stars, blue circles, and red triangles, respectively.

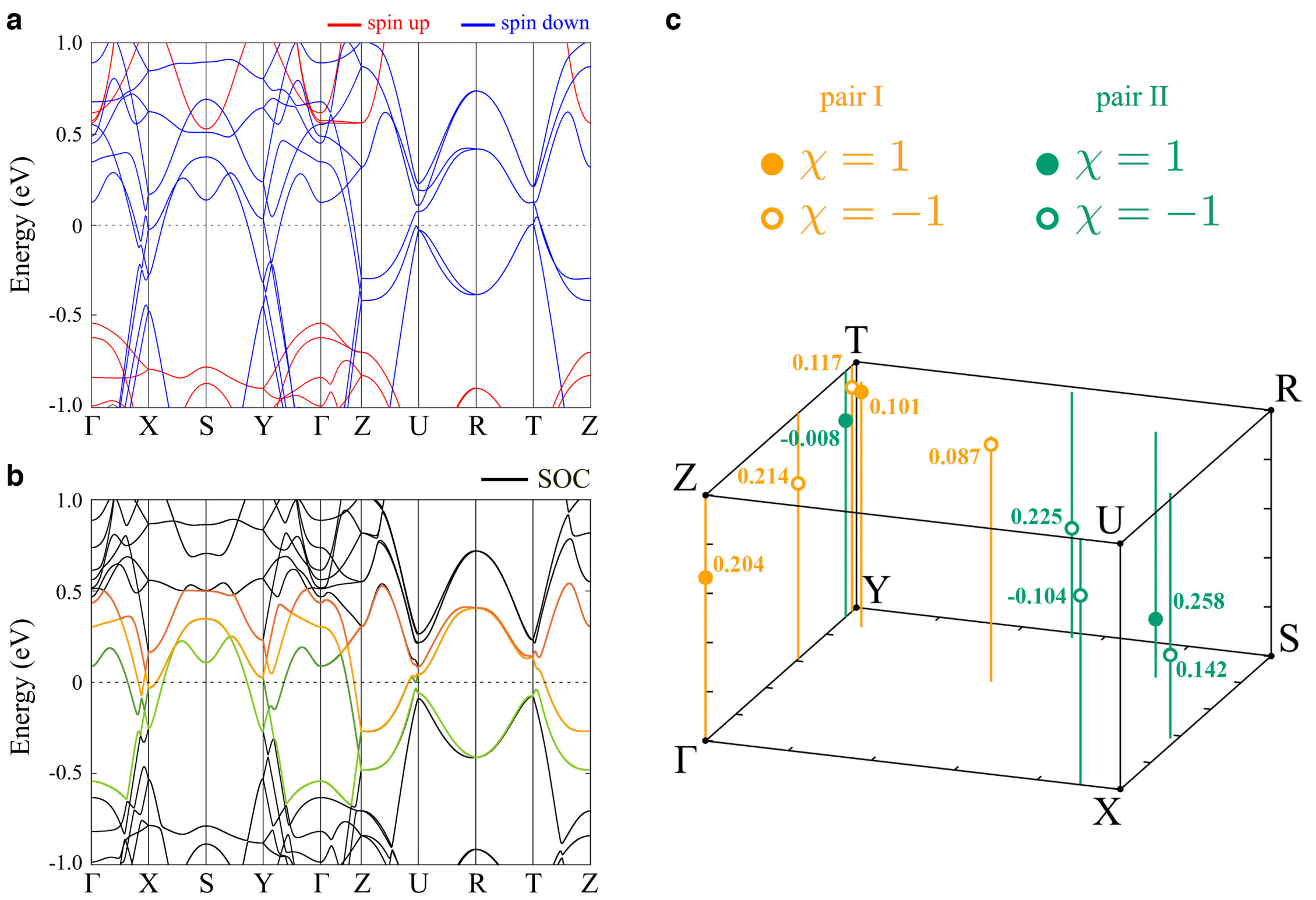

**Fig. 5 Predicted Weyl points in the orthorhombic *Pbnm* (*$D_{2h}$*) phase of $SrRuO_3$. a**, Band structure for the ferromagnetic ground state without spin-orbit coupling as obtained from GGA+$U$ calculations with $U$ = 2.6 eV and $J$ = 0.6 eV. **b**, Band structure for the ferromagnetic state with spin-orbit coupling and the magnetization along the orthorhombic $c$ axis obtained from GGA+$U$+SOC calculations with $U$ = 2.6 eV and $J$ = 0.6 eV. Two pairs of bands I (orange and dark orange bands) and II (green and dark green bands) are selected for calculating the corresponding chirality $\chi$ at each band crossing point. **c**, Positions of the Weyl points in the irreducible part of the Brillouin zone calculated for two pairs of bands I (orange filled and open circles for $\chi$ = 1 and -1, respectively) and II (green filled and open circles for $\chi$ = 1 and -1, respectively). Numbers next to the points indicate their energy distance from the Fermi level ($E$-$E_F$, where $E_F$ is the Fermi energy) in eV unit. Vertical orange and green lines indicate the in-plane (Γ-X-S-Y plane) positions of the Weyl points for easy viewing. Note that the orange and dark green bands are degenerate at the border of the Brillouin zone (X-S-Y and Z-U-R-T-Z).

## METHODS

### Magnetotransport measurements.

We first deposited the Ag electrodes on a $SrRuO_3$ surface. Then, we patterned the samples into 200 × 350 μm$^2$ Hall bar structures by photolithography and Ar ion milling. Resistivity was measured using the four-probe method at 100 μA in a Physical Property Measurement System (PPMS) DynaCool sample chamber equipped with a rotating sample stage. Low-noise measurements below 1 K were performed by an AC analog lock-in technique, and the sample was cooled down in a $^3$He-$^4$He dilution refrigerator.

### Determination of the RRR in $SrRuO_3$

The RRR was determined as the ratio of the longitudinal resistivity $\rho_{xx}$ at 300 K [$\rho_{xx}$(300 K)] and $T\rightarrow0$ [$\rho_{xx}(T\rightarrow0$ K)]. $SrRuO_3$ exhibits Fermi liquid behaviour at a low temperature, which is characterized by a linear relationship between $\rho_{xx}$ and $T^2$.[12,42] Based on this relationship, we estimated $\rho_{xx}(T\rightarrow0$ K) by extrapolating the $\rho_{xx}$ vs $T^2$ fitting line below 10 K (Extended Data Fig. 1b).

### Machine-learning-assisted MBE

We grew the high-quality $SrRuO_3$ films (63-nm thick) on (001) $SrTiO_3$ substrates in a custom-designed MBE setup equipped with multiple e-beam evaporators for Sr and Ru (Extended Data Fig. 2a). We precisely controlled the elemental fluxes, even those of elements with high melting points, *e.g.*, Ru (2250°C), by monitoring the flux rates with an electron-impact-emission-spectroscopy sensor and feeding the results back to the power supplies for the e-beam evaporators. The oxidation during growth was carried out with ozone ($O_3$) gas. $O_3$ gas (~15% $O_3$ + 85% $O_2$) was introduced through an alumina nozzle pointed at the substrate. Further information about the MBE setup and preparation of the substrates is described elsewhere[60]. The surface morphology of our $SrRuO_3$ films is composed of atomically flat terraces and steps, as observed by atomic force microscopy[28]. Together with Laue fringes in the $\theta$-$2\theta$ X-ray diffraction patterns[28], this indicates the high crystalline order and large coherent volume of our $SrRuO_3$ films.

Fine-tuning of growth conditions is essential but challenging for high-RRR $SrRuO_3$ growth. Therefore, only a few papers have reported $SrRuO_3$ films with RRRs over 50.[28,43,44] While a conventional trial-and-error approach may be a way to optimize the growth conditions, this is time-consuming as well as costly, and the optimization efficiency largely depends on the skill and experience of individual researchers. To avoid such a time-consuming approach and reduce experimental time and cost, we employed machine-learning-assisted MBE, which we developed in previous research[28]. Here, three important growth parameters [Ru flux rate, growth temperature, and nozzle-to-substrate distance (Extended Data Fig. 2a)] were optimized by a Bayesian optimization (BO) algorithm, which is a sample-efficient approach for global optimization[61]. This algorithm sequentially produces the three parameter values at which a high RRR value is predicted given past trials.

Extended Data Fig. 2b shows the procedure for machine-learning-assisted MBE growth of the $SrRuO_3$ thin films based on the BO algorithm. Here, we optimized each parameter in turn using BO. We first choose one of the growth parameters to update and fixed the rest, ran the BO algorithm to search the growth parameter, and then switched to another growth parameter. This is because BO can be inefficient in large dimensions due to the difficulty of predicting the outcome value for unseen parameters. Here, RRR = S($x$) is the target function specific to our $SrRuO_3$ films, and $x$ is the growth parameter (Ru flux rate, growth temperature, or nozzle-to-substrate distance). BO constructs a model to predict the value of S($x$) for unseen $x$ using the result of past $M$ trials $\{x_m, \mathrm{RRR}_m\}_{m=1}^{M}$, where $x_m$ are the $m$th growth parameters and $\mathrm{RRR}_m$ are the corresponding RRR values. Specifically, we use the Gaussian process regression (GPR) as a prediction model[28,62]. GPR predicts S($x$) as a Gaussian random variable following N($\mu$, $\sigma^2$). This means $\mu$ and $\sigma^2$ are calculated from $x$ and the $M$ data points. Subsequently, we choose the growth parameter $x$ in the next run such that the expected improvement (EI)[63] is maximized. EI balances the exploitation and exploration by using the predicted $\mu$ and $\sigma^2$ at $x$. This measures the expectation of improvement over the best experimental RRR so far. This routine is iterated until further improvement is no longer expected. In practice, we terminate the iteration when the number of trials reaches the predetermined budget. Here, we stopped the routine at 11 samplings per parameter. After completing 11 samplings for a certain parameter, we chose the value that gave the highest RRR and started the optimization of another parameter. In this optimization procedure, we used RRR($T$ = 4 K) instead of RRR($T$→0 K) for easy estimation. Further details about the implementation of machine-learning-assisted MBE are described elsewhere[28].

In our previous study[28], we carried out the optimization of the Ru flux rate while keeping the other parameters unaltered. Subsequently, we tuned the growth temperature and the nozzle-to-substrate distance. As a result, we obtained a high-RRR (51.8) $SrRuO_3$ film in only 24 MBE growth runs (Extended Data Fig 2c). Since the RRR was still lower than the highest value reported (~80) in the literature[12,44], we further carried out the re-optimization of the Ru flux rate and growth temperature with previously optimized parameters (the Ru flux = 4.2 Å/s, growth temperature = 721°C, and nozzle-to-substrate distance = 15 mm) as a starting point to find the global-best point in the three-dimensional parameter space. Extended Data Fig. 2c shows the highest experimental RRR values plotted as a function of the total number of MBE growth runs. With the re-optimization of the Ru flux rate and growth temperature, the highest RRR($T$ = 4 K) value increased and reached 81 in only 44 MBE growth runs. The highest experimental RRR($T$→0 K), 84.3, was achieved at the Ru flux = 3.65 Å/s, growth temperature = 781°C, and nozzle-to-substrate distance = 15 mm. The availability of such high-quality film allowed us to probe the intrinsic transport properties of $SrRuO_3$.

**Excluding other possible origins of the positive MR in $SrRuO_3$**

Since $SrRuO_3$ has complex Fermi surfaces with both topologically trivial and non-trivial bands, the contribution from the trivial bands to the MR, such as orbital MR ($\propto B^2$)[64], anisotropic MR (AMR) ($\propto$ relative angle between the electric current and the magnetization), and weak anti-localization (WAL) ($\propto B^{-1/2}$)[65], will appear in the MR in our $SrRuO_3$ films. In fact, the AMR feature is clearly observed in the near-zero field region for the $SrRuO_3$ film with the RRR of 84.3 as shown in Extended Data Fig. 3a. While it is difficult to distinguish each contribution on the MR in a low-field region (< 0.5 T), the MR in a high field region (> 0.5 T) is clearly dominated by the unsaturated linear positive MR. The MR curve at 2 K for the $SrRuO_3$ film with the RRR of 84.3 with $B$ applied in the out-of-plane [001] direction of the $SrTiO_3$ substrate (Extended Data Fig. 3b) is linear above 0.5 T. This means that the magnetic field dependences of the WAL ($\propto B^{-1/2}$) and of the orbital MR ($\propto B^2$) provide only a very limited contribution to the net MR response and that the WAL and the orbital MR are negligibly small above 0.5 T. In addition, the contribution of the AMR above 0.5 T is also negligible since the magnetization of the $SrRuO_3$ films saturates at about 0.5 T (see Extended Data Fig. 3c).

Next, the linear positive MR caused by carrier fluctuations[39,66,67,68,69,70,71], originating from disorders and/or non-uniformity of dopants, can be excluded in the $SrRuO_3$ film with the RRR of 84.3, since its Hall resistivity is not a linear function of $B$ as shown in Extended Data Fig. 3d. The carrier fluctuations can cause the linear positive MR by an admixture of a component of the Hall resistivity to the longitudinal resistivity since the mobility fluctuation makes the Hall voltage contribute to the longitudinal voltage. In ref. 39, the admixture of a component of the Hall resistivity to the longitudinal resistivity $\rho_{xx}$ due to the carrier fluctuations is expressed as:

$$\rho_{xx} \propto \frac{d\rho_{xy}}{dB} \times |B| \quad (2)$$

where $\rho_{xy}$ is the Hall resistivity, and $B$ is the external magnetic field. Therefore, $\rho_{xx}$ is proportional to $B$ when $d\rho_{xy}/dB$ is constant. For example, the GaAs 2DEG system shows clear linear MR reflecting its high-linearity of $\rho_{xy}$ in the whole measurement range (0-33 T)[39]. In our case, however, the Hall resistivity shows non-linear behaviour reflecting its semi-metallic feature (Extended Data Fig. 3d). In fact, the $d\rho_{xy}/dB \times |B|$ curve calculated by the $\rho_{xy}$ curve (Extended Data Fig. 3e) does not agree with the measured MR $(\rho_{xx}(B)-\rho_{xx}(0\ \mathrm{T}))/\rho_{xx}(0\ \mathrm{T})$ in the $SrRuO_3$ film, especially in the low-magnetic-field range (0 T < $B$ < 4 T), where the non-linearity of the Hall resistivity (Extended Data Fig. 3d) is prominent. This means that carrier fluctuations are not the origin of the linear positive MR in our $SrRuO_3$ films.

The admixture of a component of the Hall resistivity to the longitudinal resistivity is also ruled out by taking the small Hall resistivity of $SrRuO_3$ into consideration. For the admixture of the Hall voltage, to cause the large positive MR over 100%, a large Hall voltage due to a small carrier density is necessary; only semiconducting materials can show this type of large MR. For example, the GaAs 2DEG system with the small sheet carrier density of $3\times10^{11}$ $cm^{-2}$ shows by far larger linear MR (~ $10^5$ %)[39] than that of $Bi_2Se_3$ (~ 1%) with the large sheet carrier density of $1.7\times10^{16}$ $cm^{-2}$.[68] In contrast, since the $SrRuO_3$ film with the RRR of 84.3 is metallic, the Hall resistivity (= 0.36 μΩ·cm) at 2 K and 14 T is more than ten times smaller than the longitudinal resistivity (= 3.79 μΩ·cm) at 2 K and 14 T. This means that the longitudinal voltage in $SrRuO_3$ is less affected by the Hall voltage.

Therefore, we concluded that the observed unsaturated linear positive MR above 0.5 T originates from the linear energy dispersion around the Weyl nodes and that the contribution from the conventional bands on the MR is negligible in the $SrRuO_3$ film with the RRR of 84.3.

**Temperature dependence of the AHE in the $SrRuO_3$ film with the RRR = 84.3**

It is known that the AHE in $SrRuO_3$ is caused by an extrinsic factor (side-jump scattering) in addition to an intrinsic factor (KL mechanism)[12,37,38]. To determine the contributions of the intrinsic and extrinsic factors to the AHE, we investigated the temperature-dependent scaling of the AHE in the $SrRuO_3$ film with the RRR = 84.3.

The Hall resistivity $\rho_{xy}(B)$ in $SrRuO_3$ is described as the summation of the ordinary $\rho_{xy}^{\mathrm{OHE}}(B)$ and anomalous $\rho_{xy}^{\mathrm{AHE}}(B)$ components of Hall effects[12,37,38]:

$$\rho_{xy}(B) = \rho_{xy}^{\mathrm{OHE}}(B) + \rho_{xy}^{\mathrm{AHE}}(B) \quad (3)$$

$\rho_{xy}^{\mathrm{AHE}}(B)$ is proportional to the perpendicular magnetization component $M_{\perp}$: $\rho_{xy}^{\mathrm{AHE}}(B) = c\rho_s M_{\perp}$ ($c$; constant). Here, the proportional coefficient $\rho_s$ differs depending on the origin of the AHE, which can be of intrinsic (KL mechanism) and extrinsic (side jump scattering) origin[12,37,38]. As shown in Extended Data Fig. 4a, above $T_F$ (~20 K), clear AHE, which is proportional to the magnetization hysteresis curve (*e.g.*, Fig. 1d, right inset), is discernible, and it dominates Hall resistivity in a near-zero magnetic field. On the other hand, below $T_F$, the AHE components are negligibly small.

The temperature dependence of the AHE in $SrRuO_3$ has been well reproduced by a model where both the intrinsic KL mechanism and the extrinsic side-jump scattering terms are taken into account[38]. In this model, the relationship between $\rho_s$ and $\rho_{xx}$ is described as

$$\rho_s = \frac{a_1}{\Delta^2 + a_2\rho_{xx}^2}\rho_{xx}^2 + a_3\rho_{xx}^2 \tag{4}$$

where $a_1$, $a_2$, $a_3$, and $\Delta$ are the fitting parameters, which are associated with the band structure[38]. The first term describes a contribution from the off-diagonal matrix elements of the velocity operators, called the KL term in the model[38], and the second term is the contribution from the side-jump scattering. In this model, the KL term is obtained by incorporating the finite scattering rate, which is inversely proportional to $\rho_{xx}$, into Kubo's formula[72], and the constant $a_1$ in eq. (4) is expressed as

$$a_1 = b\mathrm{Im}(A), \tag{5}$$

where $b$ is a constant and $A$ is expressed as

$$A = \int_{\mathbb{K}} \frac{d^3k}{2\pi^3} \langle 1k|v_y|2k\rangle \langle 2k|v_x|1k\rangle. \tag{6}$$

Here, $A$ can be associated with a Berry phase[73]. The $v_y$ and $v_x$ are the velocity operators in the direction of $y$ and $x$ axes, respectively, $\mathbb{K}$ is a set of quasi-momentum of the states producing the dominant contribution to $A$, and $k \in \mathbb{K}$. Here, "1" and "2" are indices to denote two individual bands: the former crosses the Fermi level, while the latter is fully occupied. When the system has multiple bands crossing the $E_F$, the $a_1$ value is expressed by the sum of the $A$ values for each band. Therefore, the relationship between $\rho_s$ and $\rho_{xx}$ (eq. (4)) does not even change in the system with multiple bands. Accordingly, this model can be applied to analyse the temperature dependence of the AHE of $SrRuO_3$, although detailed information about the electronic bands near $E_F$ in $SrRuO_3$ is required to assign the contribution of each band to the intrinsic AHE.

Extended Data Fig. 4b shows the $\rho_s$ vs $\rho_{xx}$ plot of the $SrRuO_3$ film with the RRR = 84.3 and the fitting result of eq. (4). The fitting curve reproduces the AHE sufficiently, indicating that the AHE in the $SrRuO_3$ film arises from the intrinsic KL mechanisms along with the extrinsic side-jump scattering. The important point in eq. (4) is that the AHE is asymptotic to zero when $\rho_{xx}\rightarrow 0$. Accordingly, for the $SrRuO_3$ film with such a very high RRR (84.3), equivalently with a very small residual $\rho_{xx}$, AHE becomes negligibly small at low temperatures. Therefore, $\rho_{xy}(B)$ curves below $T_F$ in our data are dominated by the ordinary Hall effect.

**Excluding other possible origins of the negative MR in $SrRuO_3$**

Here, we exclude other possible mechanisms as the origin of the anisotropic linear negative MR at 2 K for the $SrRuO_3$ film with the RRR of 84.3. As described in the Methods section "Excluding other possible origins of the positive MR in $SrRuO_3$", the contribution from the trivial bands to the MR, such as orbital MR[64], AMR, and WAL[65], are negligible above 0.5 T.

The electron-magnon scattering[74] is also excluded as an origin of the linear negative MR (Fig. 2). The measurement temperature (2 K) of the anisotropic linear negative MR is too low to excite magnons in $SrRuO_3$. A recent inelastic neutron scattering study on single-crystalline $SrRuO_3$[75] reveals that the magnon gap of $SrRuO_3$ is equal to 11.5 K. Therefore, the contribution from magnon scattering to the negative MR is negligible.

Finally, the boundary scattering effect is also excluded. The values of mean free paths $l_m$ for the Weyl fermions ($F_1$ (= 26 T) and $F_2$ (= 44 T) orbits), estimated from the SdH oscillation in Fig. 3b, are 89 and 132 nm, respectively. Here, the $l_m$ values are estimated by $l_m = \hbar v_F/(2\pi k_B T_D)$ (where $v_F$ is Fermi velocity, $k_B$ is Boltzmann's constant, and $T_D$ is the Dingle temperature). The mean free paths of other orbits ($F_3$, $F_4$, $F_5$, and $F_6$) are smaller than these two orbits ($F_1$ and $F_2$). Since the mean free paths of the Weyl fermions are larger than the thickness of the $SrRuO_3$ films ($d_{SrRuO3}$ = 63 nm), the boundary scattering may contribute to the negative MR. However, we can exclude its contribution as described as follows. The boundary-scattering-induced MR should saturate in a high-magnetic-field region[76], since boundary scattering does not occur when cyclotron diameter $d_c$ (= $\hbar k_F/(\pi e B)$; $k_F$, Fermi wave number) is smaller than the film thickness ($d_{SrRuO3}$ = 63 nm). If the MR had been induced by boundary scattering, it would have saturated when $B$ is larger than 0.93 and 1.2 T for the $F_1$ and $F_2$ orbits, respectively, at which $d_c$ becomes comparable to $d_{SrRuO3}$. On the contrary, the observed negative MR (Fig. 2a and c) does not show saturation behaviour with $B$. Note that the coherent lengths of the other orbits ($F_3$, $F_4$, $F_5$, and $F_6$) are not long enough to travel a full cyclotron trajectory at the measurement temperature (= 2 K), where only the $F_1$ and $F_2$ SdH oscillation peaks are detected; as shown in Fig. 3d, the SdH oscillation peaks of $F_3$, $F_4$, $F_5$, and $F_6$ are observed at 0.1K. Therefore, the contribution of the boundary scattering to the negative MR (Fig. 2a and c) is negligibly small.

**Effect of the Fermi-level pinning on the phase shift in the SdH oscillation**

A large number of carriers in 3D multiband systems pin the chemical potential. In 2018, Kuntsevich *et al.* phenomenologically explained that, even when a large number of carriers in normal bands pin the chemical potential, the SdH oscillation of the Dirac (linear) dispersion should show the π phase shift[77]. Here, we will explain why the chemical potential pinning does not affect the phase shift in the SdH oscillations, along with ref. 77.

We will begin with the general explanation of SdH oscillations without the Fermi-level (chemical potential) pinning. In the SdH oscillation, the minimal conductivity is obtained when the chemical potential $\mu_c$ is located in the mid-gap between the Landau levels (Extended Data Fig. 5a). For this $\mu_c$, the $N$th Landau level beneath the chemical potential is fully occupied and the density of states shows the minimal value. Reflecting this situation, in the LK theory[87], the magnetic field at which the magnitude of the SdH oscillation takes a minimal value is expressed as

$$\frac{F_i}{B_N^{\min}} = N - \beta_{Bi}. \quad (7)$$

Here, $F_i$ is the frequency of the SdH oscillation, $B_N^{min}$ is the magnetic field giving the minimal value of the conductivity at the $N$th Landau level, and $2\pi\beta_{Bi}$ is the phase shift. The important point of eq. (7) is that the $1/B_N^{\min}$ can be expressed as a linear function of $N$. This ensures the validity of using eq. (7) for the determination of the phase shift caused by the Berry phase as we did in the main manuscript.

Next, we consider the situation where a large number of carriers in normal bands pin the chemical potential, i.e. $\mu_c(B) = const.$ The $N$th Landau levels of the quadratic energy band $\epsilon_{N,Q}$ and the linear energy band $\epsilon_{N,D}$ are expressed as[83,77]

$$\epsilon_{N,Q} = \frac{\hbar eB}{m}\left(N + \frac{1}{2}\right) \quad (8)$$

and

$$\epsilon_{N,D} = \sqrt{2N\hbar eBv^2}, \quad (9)$$

respectively, where $m$ is an effective mass, $v$ is the velocity of electrons in the linear band, and $N$ is the Landau level index. The conductivity becomes maximum at $B$ with which the Landau level crosses $\mu_c$, since the density at $\mu_c$ becomes maximum when $\mu_c$ is located at the center of the Landau level (Extended Data Fig. 5b). Therefore, by solving the equation $\epsilon_{N,Q(D)} = \mu_c$, we can obtain the relationships between $N$ and the magnetic field $B_{N,Q(D)}^{\max}$ at which the conductivity becomes maximum as

$$\hbar\frac{eB_{N,Q}^{max}}{m}\left(N + \frac{1}{2}\right) = \mu_c = const. \quad (10)$$

and

$$2N\hbar eB_{N,D}^{max}v^2 = \mu_c^2 = const., \quad (11)$$

for the quadratic and linear bands, respectively. Since $B_{N,Q(D)}^{min}$ is located roughly at the middle point between $B_{N-1,Q(D)}^{\max}$ and $B_{NQ(D)}^{\max}$, we can obtain the relationship between $B_{NQ(D)}^{\min}$ and $N$ by shifting $N$ in eqs. (10) and (11) by a half-integer as

$$\frac{\mu_c m}{e\hbar}\frac{1}{B_{N,Q}^{\min}} = N \quad (12)$$

and

$$\frac{\mu_c^2}{2\hbar ev^2}\frac{1}{B_{N,D}^{\min}} = N - \frac{1}{2} \quad (13)$$

respectively. Then, by using the relationship between the $\mu_c$ and the Fermi wave number $k_F$ for the quadratic and linear band, $\mu_c = (\hbar k_F)^2/2m$ and $\mu_c = \hbar v k_F$, respectively, and the relationship between the Fermi surface area $S$ (= $\pi k_F^2$) and $F_i$, $S = 2\pi eF_i/\hbar$, eqs. (12) and (13) can be simply expressed as

$$\frac{F_i}{B_{N,Q}^{\min}} = N \quad (14)$$

and

$$\frac{F_i}{B_{N,D}^{\min}} = N - \frac{1}{2}. \quad (15)$$

Eqs. (14) and (15) mean that the trivial (quadratic) and non-trivial (linear) dispersion provides 0 and $\pi$ (= $2\pi \times 1/2$) phase shifts, respectively, even when a large number of carriers pin the chemical potential. Therefore, whether the chemical potential is pinned or not in our $SrRuO_3$ film, we can estimate the phase shift by using the LK theory [eq. (1)]. Physically, the difference in the phase shifts in eq. (14) and (15) comes from the difference in the Berry phases of quadratic and linear dispersions[77].

In a 3D system with $B$ applied in the $k_z$ direction, a two-dimensional cyclotron motion occurs in the $k_x$-$k_y$ plane in 3D $k$-space at the $k_z$ position where the area of the Fermi surface takes an extremal value[80]. This cyclotron orbit is called an "extremal orbit". As in the case of 2DEG Rashba systems, graphene, and topological surface states, observation of the π Berry phase, which originates from a band touching point of the Weyl node and accumulates along the extremal orbit, is one of the important signatures of Weyl fermions[20,23,24]. Therefore, we think that the observed π Berry phase in the SdH oscillations is acquired by a surface integral of the Berry curvature Ω over a closed surface containing a Weyl point in $k$-space (Fig. 3a).

**Data pretreatment for quantum oscillations**

Pretreatments of the SdH oscillation data are crucial for deconvoluting quantum oscillation spectra, since magnetoconductivity data generally contain not only oscillation components but also other magnetoresistive components as background signals[43]. In particular, SdH oscillations in $SrRuO_3$ are subject to being masked by large non-saturated positive MR (Extended Data Fig. 5c)[43]. Here, we subtracted the background using a polynomial function up to the fifth order and extracted the oscillation components as shown in Extended Data Fig. 5d. Then, we carried out the well-established pretreatment procedure for Fourier transform of quantum oscillations[78,79]. First, we interpolated the background-subtracted data to prepare an equally spaced data set as a function of $1/B$. Then, we multiplied the Hanning window function to obtain the periodicity of the experimental data. Finally, we performed fast Fourier transform on the treated data set.

**Exclusion of other possible mechanisms of the phase shift in $SrRuO_3$**

Here, we exclude other possible non-topological effects that could cause the phase shift in SdH oscillations, which are the mosaic effect, magnetic breakdown, and Zeeman splitting.

The phase shift and deviation of the SdH oscillations from the conventional LK theory, which occur at crystal grains and magnetic domains, are collectively called the mosaic effect[80,81]. This effect may occur in samples having multiple crystal domains, such as polycrystals, or in ferromagnets having magnetic domain structures. However, it should be negligible in our samples, because they are high-quality single-crystalline thin films as shown by a STEM image (Fig. 1c, Extended Data Fig. 1a, and Extended Data Fig. 5e) and they are free from magnetic domain structures as shown in Fig. R2b where magnetization is saturated at about 0.5 T (Extended Data Fig. 3c).

Next, the possibility of magnetic breakdown is also ruled out since there is no magnetic breakdown orbits in the SdH oscillations for the $SrRuO_3$ film with the RRR of 84.3 (Fig. 3). Magnetic breakdown occurs when different orbits approach each other closely in $k$-space under the presence of large magnetic fields, resulting in a new orbit (a magnetic breakdown orbit) whose frequency is given by the sum of the frequencies of the original orbits[82]. If magnetic breakdown had occurred in our measurement field range ($|B|$ < 14 T), we would have observed magnetic breakdown orbits whose frequencies are $F_1$

+ $F_2$ (= 70 T), $F_1$ + $F_3$ (= 326 T), and so on. However, we did not find such frequencies in our quantum oscillation analysis as shown in Extended Data Table 1. In particular, $F_1$ and $F_2$, which are responsible for the non-trivial phase shifts, cannot be produced by the sum of the other orbitals' frequencies. Therefore, magnetic breakdown is not the origin of the phase shift.

Finally, we focus on the effect of Zeeman splitting on the phase shift in a magnetic Weyl semimetal. The condition where the quantum oscillation takes minimal or maximum values is expressed as[83]

$$\frac{F}{H} = n + \gamma \pm \frac{1}{2}S. \quad (16)$$

Here, $F$ is a frequency, $H$ is a magnetic field, $n$ is an integer value, γ indicates the phase shift caused by the Berry phase, and $S$ is the spin-splitting parameter from Zeeman effect. The ± sign indicates the up/down spins in each Landau level. Due to the Zeeman effect, every Landau level is split-off by the magnetic field, and finally it affects the phase shift γ through the change of the $S$ value. In fact, this Zeeman splitting of Landau levels changes SdH spectra in Weyl semimetals and magnetic Dirac materials[84,85], in which Landau levels are degenerate, and this effect has to be taken into account to deduce the phase shift γ from the experimental data. By contrast, in $SrRuO_3$, all Landau levels are nondegenerate since the ferromagnetic exchange coupling lifts the spin degeneracy of all the electronic bands crossing the $E_F$[12,42,43,86]. Altogether, since the $S$ value of eq. (16) is zero in $SrRuO_3$, we can simply estimate the phase shift γ by the LK theory[87], in which $S$ is not taken into account, and assign the phase shift in the SdH oscillations to the Berry phase accumulation along the cyclotron orbits.

**SdH oscillations of trivial orbits**

Together with SdH oscillations from the non-trivial orbits having low frequencies ($F_1$ and $F_2$) (Fig. 3b, c), we observed SdH oscillations from the trivial orbits having high frequencies ($F_3$, $F_4$, $F_5$, and $F_6$) at 0.07 K < $T$ < 0.75 K and 12.5 T < $B$ < 14 T (Fig. 3d and Extended Data Fig. 6b). Since the carriers in the trivial orbits in $SrRuO_3$ are expected to have larger effective masses than those in the $F_1$ and $F_2$ orbits[12,43,58], measurements of the former oscillations should be carried out in relatively low-$T$ and high-$B$ regions. We estimated the cyclotron masses of the carriers in $F_3$, $F_4$, $F_5$, and $F_6$ orbits from the temperature dependences of the respective peaks based on the LK theory (Extended Data Fig. 6c-f). In the LK theory for the mass estimation, $B$ is determined as the interval value in the magnetic field range. Extended Data Table 1 shows the estimated cyclotron masses for $F_1$, $F_2$, $F_3$, $F_4$, $F_5$, and $F_6$. The cyclotron masses in the $F_3$, $F_4$, $F_5$, and $F_6$ orbits are relatively high (> $2.8m_0$), reflecting the trivial band structure (energy dispersions) as their origin.

**RRR dependence of the ferromagnetism, Fermi liquid behaviour, and Weyl behaviour**

In Fig. 4a-c, we investigated the RRR dependence of the ferromagnetism, Fermi liquid behaviour, and Weyl behaviour in $SrRuO_3$. For the ferromagnetism, $T_C$ values are estimated as the position of the kinks in $\rho_{xx}$ vs $T$ curves as shown in Extended Data Fig. 7a. For the Fermi liquid behaviour, we defined the Fermi liquid region ($T < T_F$) as the temperature range where the experimental $\rho_{xx}$ and the linear fitting line in $\rho_{xx}$ vs $T^2$ are close enough to each other (< 0.1 μΩ cm) as shown in Extended Data Fig. 7b. The upper limit temperature for measuring Weyl behaviour in $SrRuO_3$ ($T_W$) is defined as the highest temperature at which the resistivity at zero field is lower than that at 9 T ($\rho_{xx}$(0 T) < $\rho_{xx}$(9 T)).

**RRR dependence of the Hall resistivity**

Extended Data Figure 7c shows the Hall resistivity $\rho_{xy}(B)$ curves of the different RRR samples at 2 or 2.3 K with $B$ applied in the out-of-plane [001] direction of the $SrTiO_3$ substrate. As we explained in the main manuscript, the $\rho_{xy}(B)$ curves of the $SrRuO_3$ film with the RRR of 84.3 at 2 K is nonlinear, indicating the coexistence of multiple types of the Weyl fermions from which the unsaturated linear positive MR stems. Notably, as shown in Extended Data Fig. 7c, $d\rho_{xy}/dB$ changes its sign from negative to positive with decreasing RRR. In the $SrRuO_3$ film with the RRR of 8.93, clear AHE is observed near zero magnetic field due to its large residual resistivity of 20.2 μΩ·cm, and the $\rho_{xy}(B)$ curve shows the linear dependence on $B$ above 5 T as highlighted in Extended Data Fig. 7c. The carrier concentration and the mobility of the holes of the $SrRuO_3$ film with the RRR of 8.93, which are estimated from the slope of the $\rho_{xy}(B)$ above 5 T, are $4.04\times10^{22}$ $cm^{-3}$ and 7.65 $cm^2$/Vs, respectively. The carrier concentration and the mobility are consistent with the reported values for the trivial Ru 4$d$ bands crossing the $E_F$ in $SrRuO_3$[44,88]. These results mean that the Weyl fermions become more dominant in the transport properties with increasing RRR and that the contribution of the Weyl fermions on the Hall resistivity is negligibly small when the RRR is 8.93.

As described in the manuscript, the unsaturated linear positive MR also becomes more prominent with increasing RRR and decreasing temperature below $T_F$, indicating again that the non-linear $\rho_{xy}(B)$ is a hallmark of the existence of the Weyl fermions in $SrRuO_3$ and that the Weyl fermions become dominant in the transport when scatterings from impurities and phonons are sufficiently suppressed.

**Computational details**

Electronic structure calculations were performed within the density functional theory and generalized gradient approximation (GGA, Perdew-Burke-Ernzerhof)[89] for the exchange correlation functional in the projector-augmented plane wave (PAW) formalism[90] as implemented in the Vienna ab-initio Simulation package[91] (VASP). The energy cutoff was set to 500 eV, the Brillouin zone was sampled by an 8×8×6 Monkhorst-Pack mesh[92], and the convergence criterion for the electronic density was defined as $10^{-8}$ eV. The effect of electronic correlations in the Ru $d$ shell (4$d^4$ for $Ru^{4+}$) was taken into account by using the rotationally invariant GGA+$U$ scheme[93] with $U$ = 2.6 eV and $J$ = 0.6 eV. The choice of parameters is justified by early estimations[94] and is in agreement with other studies[95,96].

**First-principles calculations of Weyl points**

The orthorhombic phase of $SrRuO_3$ has the *Pbnm* (#62) space group, which corresponds to the $D_{2h}$ point group with symmetries of inversion ($I$), three mirror planes ($m_x$, $m_y$, $m_z$), and 180° rotations around the orthorhombic axes ($C_x$, $C_y$, $C_z$). The crystal structure parameters considered in the present study are $a$ = 5.5670 Å, $b$ = 5.5304 Å, $c$ = 7.8446 Å,[12] and the atomic Wyckoff positions in fractional coordinates are given as 4$c$ (0.5027, 0.5157, 0.25) for Sr, 4$b$ (0.25, 0, 0) for Ru, and 8$d$ (0.7248, 0.2764, 0.0278) for O.

The results of electronic structure calculations with and without spin-orbit coupling (SOC) are shown in Extended Data Fig. 8. One can see that the bands close to the Fermi level are formed by the Ru $4d$ states hybridized with the O $2p$ states and the electronic spectrum reveals half-metallicity (Extended Data Fig. 8a) in agreement with previous electronic structure calculations[59]. The ferromagnetic alignment was found to be the ground state configuration with the easy axis along the orthorhombic $b$ axis ($E_{010} - E_{001}$ = -2.17 meV/f.u. and $E_{010} - E_{100}$ = -0.35 meV/f.u.), and the calculated magnetic moments per Ru ion (Extended Data Table 2) are close to the experimental saturation magnetic moment of 1.25 $\mu_B$/Ru.[28] The ferromagnetic state reduces the symmetry to the $C_{2h}$ point group with one mirror plane and one rotation axis symmetry, perpendicular and parallel to the magnetization direction, respectively.

For numerical identification of the Weyl points, one needs to have a band structure with high resolution in the reciprocal space. To interpolate the resulting electronic spectrum, we employed maximally localized Wannier functions as implemented in the wannier90 package[97]. The wannierization was carried out by projecting the bands corresponding to the Ru $e_g$ and $t_{2g}$ states onto the atomic $d$ orbitals in the local coordinate frame at each Ru site.

To locate the points of degeneracy between the bands in the reciprocal space, we performed a steepest-descent minimization of the gap function $\Delta = (E_{n+1,\boldsymbol{k}} - E_{n,\boldsymbol{k}})^2$ on a uniform grid of up to 31×31×31 covering the full Brillouin zone, where the bands are considered degenerate when the gap is below the threshold of $10^{-5}$ eV[98]. To eliminate accidental crossings, the corresponding chirality at each identified point is determined by evaluating the outward Berry flux enclosed in a small sphere. The calculated chiralities $\chi$ should obey the following symmetry properties: $\chi$ does not change its sign under 180° rotations around the magnetization axis and changes its sign under mirror reflection and inversion. In this study, we only consider the points with $\chi = \pm 1$.

We have selected two pairs of bands I and II for the cases when the magnetization is along the orthorhombic $c$ and $b$ axes, as shown in Extended Data Fig. 9 and Extended Data Fig. 10, respectively. The corresponding gap function $\Delta \leq 0.1$ eV is calculated to demonstrate the proximity of the selected bands. From the number of the calculated band crossings, the Weyl points were identified as the ones that have close energy positions and $k$-point coordinates and whose chiralities obey the symmetry properties in the full Brillouin zone. The resulting Weyl points are listed in Extended Data Tables 3-6. Numerical differences in the coordinates of the Weyl points can be attributed to small spin canting (see Extended Data Table 2), which *slightly* breaks inversion symmetry (the reflection and rotation symmetries along to the magnetization are intact). From Extended Data Tables 3-6, most of the Weyl points are found to exist within an energy range of −0.2 – 0.2 eV around the Fermi level. In particular, $|E - E_F|$ for $WP_z6_{1\text{-}4}$ (8-16 meV), which are located near the Y-T line in the Z-Γ-Y-T plane as shown in Fig. 5c, is very close to the experimental chemical potentials of the Weyl fermions estimated from the SdH oscillations ($\mu_1$ = 8.5 meV and $\mu_2$ = 8.8 meV for the $F_1$ and $F_2$ orbitals, respectively).

From the obtained results, one can clearly see the presence of the Weyl fermions below and above the Fermi level coexisting with trivial half-metallic bands. However, it is worth commenting on another scenario. According to previous theoretical studies[95,96], there is no clear consensus on whether the electronic spectrum of orthorhombic $SrRuO_3$ in the ferromagnetic state is half-metallic or not, and the result turns out to depend on the details of electronic structure calculations. While our theoretical results are in good agreement with the present experiments, we do not rule out the possibility of a non-half-metallic behaviour with both spin channels crossing close to the Fermi level. Assuming that the spin-up states can also lie at the Fermi level, there will be an extra set of band crossings in addition to the Weyl points already defined in a half-metallic scenario.

Finally, it is worth noting that a small monoclinic distortion induced by the $SrTiO_3$ substrate breaks the orthorhombic $D_{2h}$ symmetry. The reported crystal structure parameters of epitaxial $SrRuO_3$ on $SrTiO_3$ are $a$ = 5.5290 Å, $b$ = 5.5770 Å, $c$ = 7.8100 Å, $\alpha$=89.41°.[12] According to our electronic structure calculations within GGA+$U$ with $U$ = 2.6 eV and $J$ = 0.6 eV for the monoclinic $SrRuO_3$ on $SrTiO_3$ (Extended Data Fig. 11), the electronic spectrum does not reveal any qualitative changes from that for the orthorhombic $D_{2h}$ symmetry, while additional band crossings may occur due to the lowered crystal symmetry.

**Extended Data Figures and Figure Legends**

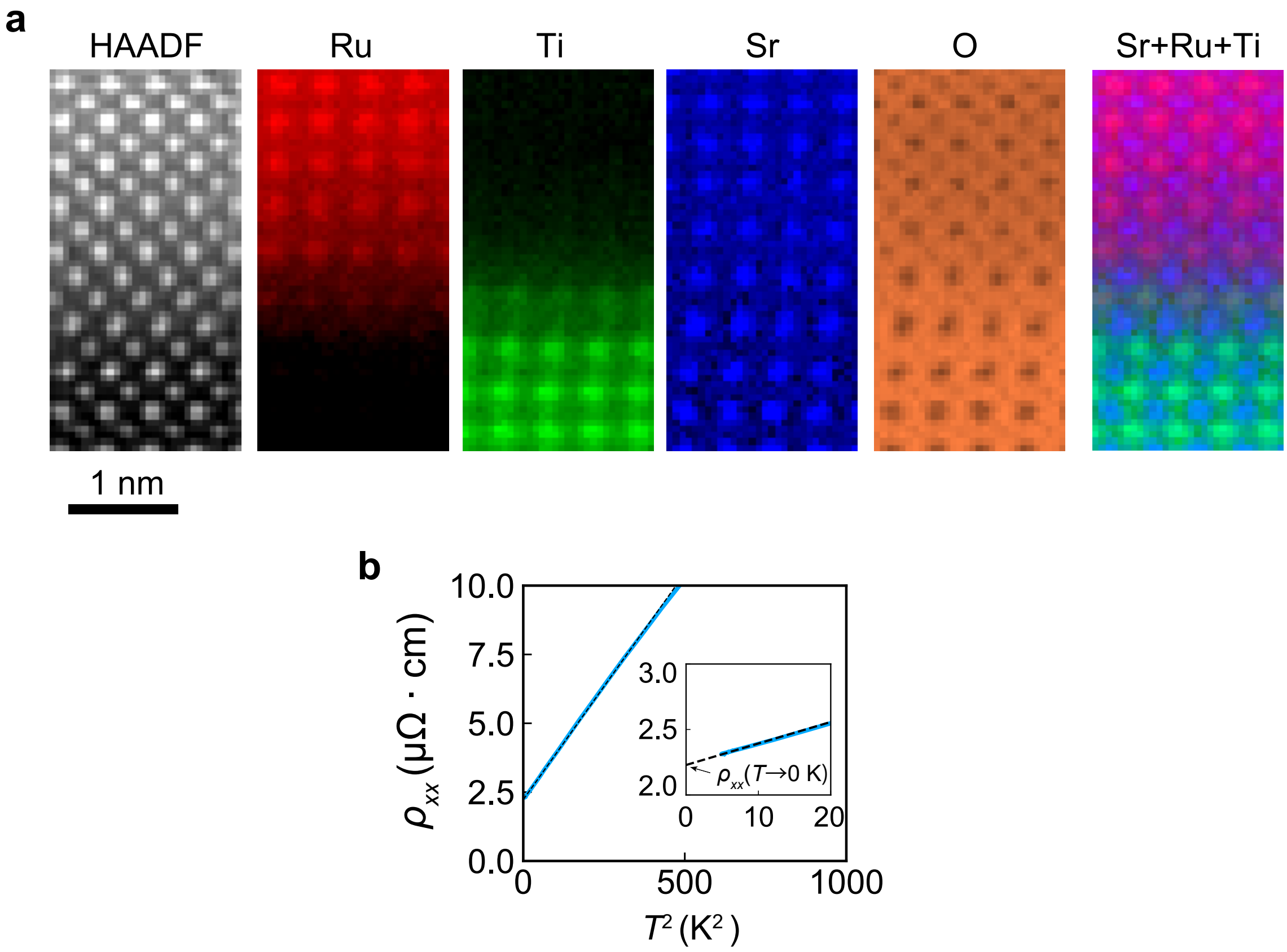

**Extended Data Fig. 1 HAADF-STEM and EELS-STEM images, and Fermi liquid behaviour in $SrRuO_3$. a,** (From left to right) HAADF-STEM image of the $SrRuO_3$ film with the RRR of 71 taken along the [100] axis of the $SrTiO_3$ substrate. EELS-STEM images for the Ru-$M_{4,5}$- (red), Ti-$L_{2,3}$- (green), Sr-$M_3$- (blue), O-$K$-edge (orange), and a color overlay of the EELS-STEM images for Sr, Ru and Ti. **b,** $\rho_{xx}(T)$ versus $T^2$ curve (blue line) for the $SrRuO_3$ film with the RRR of 84.3. The black dashed line is the linear fitting result. Close-up near 0 K$^2$ is shown in the inset. The $\rho_{xx}(T \rightarrow 0$ K) value is estimated from the extrapolation of the fitting line to the 0 K$^2$ axis.

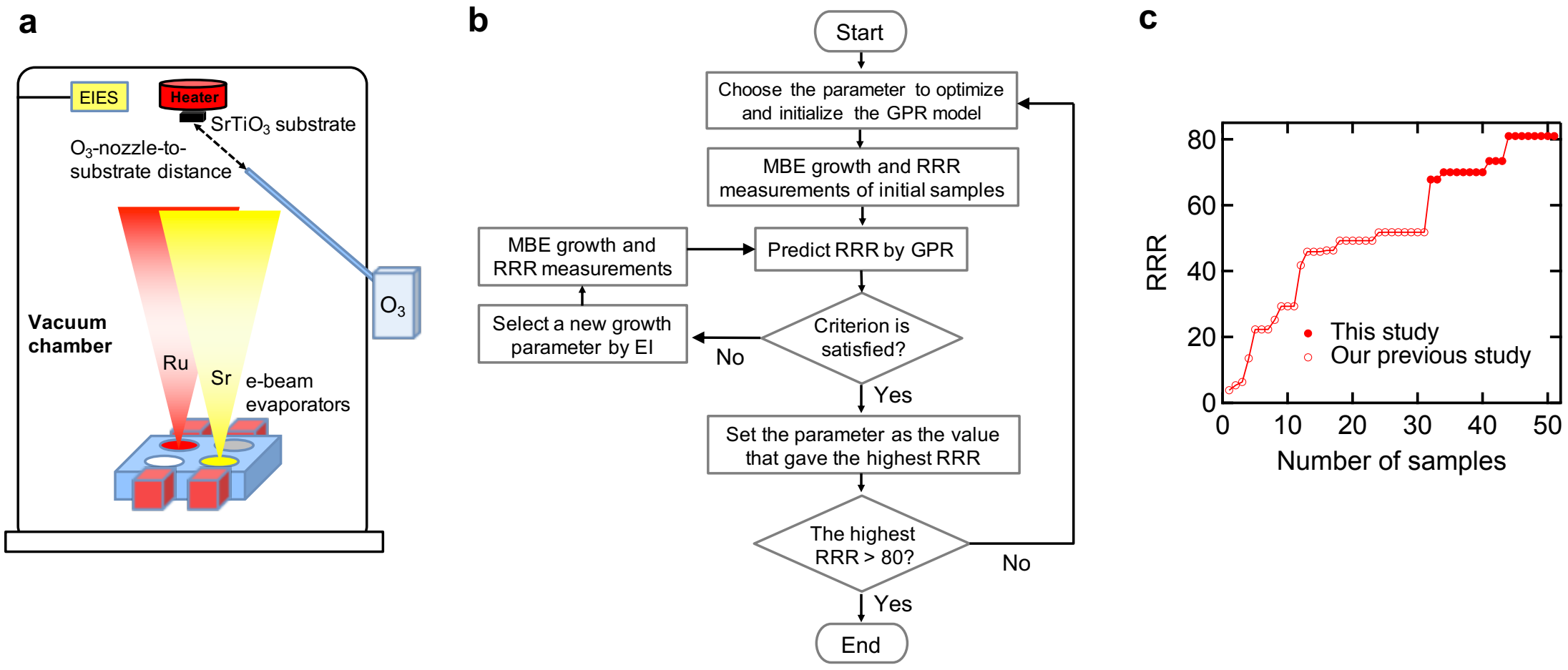

**Extended Data Fig. 2 Machine-learning-assisted MBE. a**, Schematic illustration of our multi-source oxide MBE system. EIES: electron impact emission spectroscopy. **b**, Flowchart of machine-learning-assisted MBE growth based on the BO algorithm. **c**, Highest experimental RRR values plotted as a function of the total number of MBE growth runs. In **c**, open circles are data deduced from Ref. 28. Here, the Ru flux rate, growth temperature, and nozzle-to-substrate distance were varied in ranges between 0.18 and 0.61 Å/s, 565 and 815°C, and 1 and 31 mm, in correspondence to the search ranges in BO.

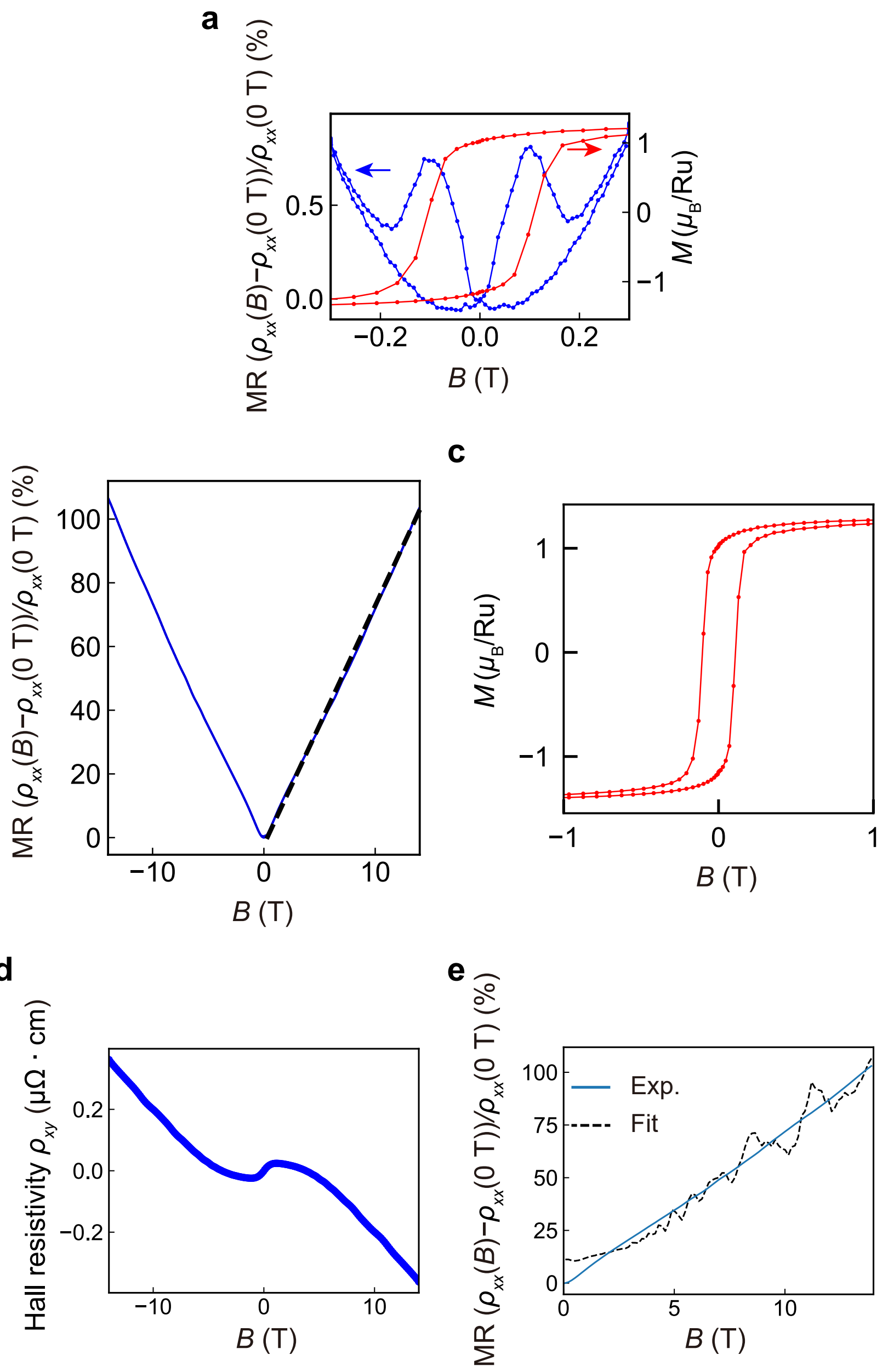

**Extended Data Fig. 3 Anisotropic MR and linear positive MR in $SrRuO_3$ a,** MR ($\rho_{xx}(B)-\rho_{xx}$(0 T))/$\rho_{xx}$(0 T) at 2 K for the $SrRuO_3$ film with the RRR of 84.3 with -0.3 T < $B$ < 0.3 T applied in the out-of-plane [001] direction of the $SrTiO_3$ substrate (blue filled circles). Magnetization $M$ versus $B$ (-0.3 T < $B$ < 0.3 T) curve at 10 K with $B$ applied in the out-of-plane [001] direction of the $SrTiO_3$ substrate (red filled circles). **b**, MR ($\rho_{xx}(B)-\rho_{xx}$(0 T))/$\rho_{xx}$(0 T) at 2 K for the $SrRuO_3$ film with the RRR of 84.3 with −14 T < $B$ < 14 T applied in the out-of-plane [001] direction of the $SrTiO_3$ substrate. The black dashed line is an eye-guide to indicate the linearity of the MR. These are the same data in Fig. 1e in the main manuscript. **c,** Magnetization $M$ versus $B$ (-1 T < $B$ < 1 T) curve at 10

K with $B$ applied in the out-of-plane [001] direction of the $SrTiO_3$ substrate. **d,** Hall resistivity $\rho_{xy}(B)$ curve at 2 K for the $SrRuO_3$ film with the RRR of 84.3 with $B$ applied in the out-of-plane [001] direction of the $SrTiO_3$ substrate. **e,** MR $(\rho_{xx}(B)-\rho_{xx}(0\ \mathrm{T}))/\rho_{xx}(0\ \mathrm{T})$ observed from 0 to 14 T at 2 K (light blue line) and the fitting result by eq. (2) (black dashed line). In **d** and **e**, $\rho_{xy}(B)$ and $\rho_{xx}(B)$ are the same data in Fig. 1f and 1e, respectively. The oscillating behaviour of the fitting curve is due to the quantum oscillation of $\rho_{xy}(B)$.

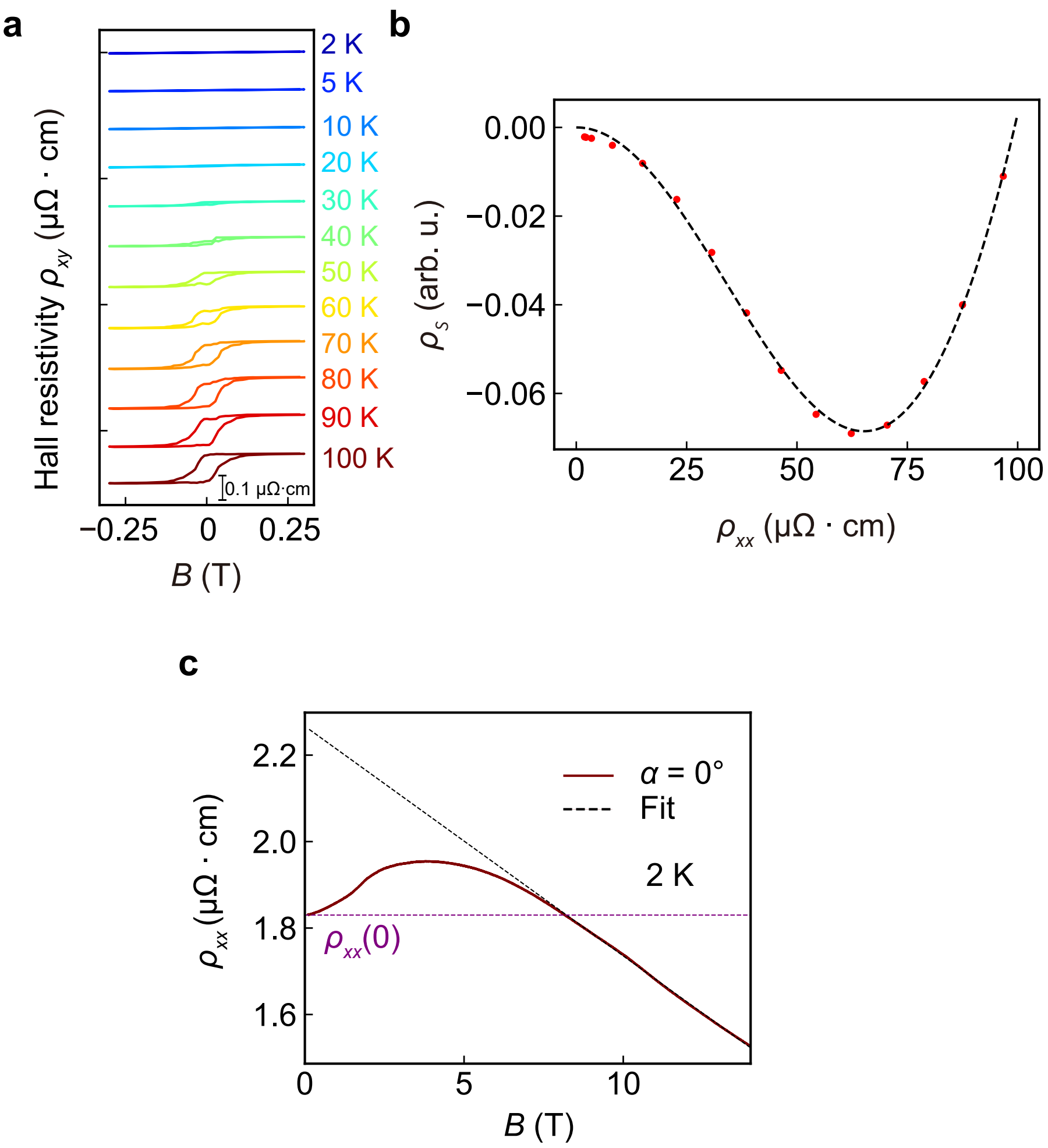

**Extended Data Fig. 4 Temperature dependence of the Hall resistivity, chiral-anomaly-induced linear and negative MR, and pretreatment of the SdH oscillation data. a,** Hall resistivity $\rho_{xy}(B)$ curves at 2 to 100 K for the $SrRuO_3$ film with the RRR of 84.3 with −0.3 T < $B$ < 0.3 T applied in the out-of-plane [001] direction of the $SrTiO_3$ substrate. In **a**, the Hall resistivity at each temperature has been offset by 0.15 μΩ·cm for easy viewing. **b,** $\rho_s$ versus $\rho_{xx}$ plot (red circles) and fitting results by eq. (4) (black dashed curve) up to 130 K. In **b**, $\rho_s$ at each temperature was obtained by dividing $\rho_{xy}$(0 T) in Fig. 1f by $M_\perp$ with 100 Oe obtained by the SQUID measurements. Here, the magnitude of $\rho_{xy}$(0 T) is defined as the averaged absolute value of $\rho_{xy}$ at ± 0.2 T. $\rho_{xx}$ at each temperature is the same data as in Fig. 1d. **c,** $\rho_{xx}(B)$ at $\alpha$ = 0 (brown solid curve) for the $SrRuO_3$ film with the RRR of 84.3 and the linear fitting line (black dashed line) to $\rho_{xx}(B)$ in the negative MR region (8 T < $B$ < 14 T). The fitting line completely reproduces the negative and linear MR region. The purple dashed line corresponds to the value of the zero field resistivity $\rho_{xx}$(0 T).

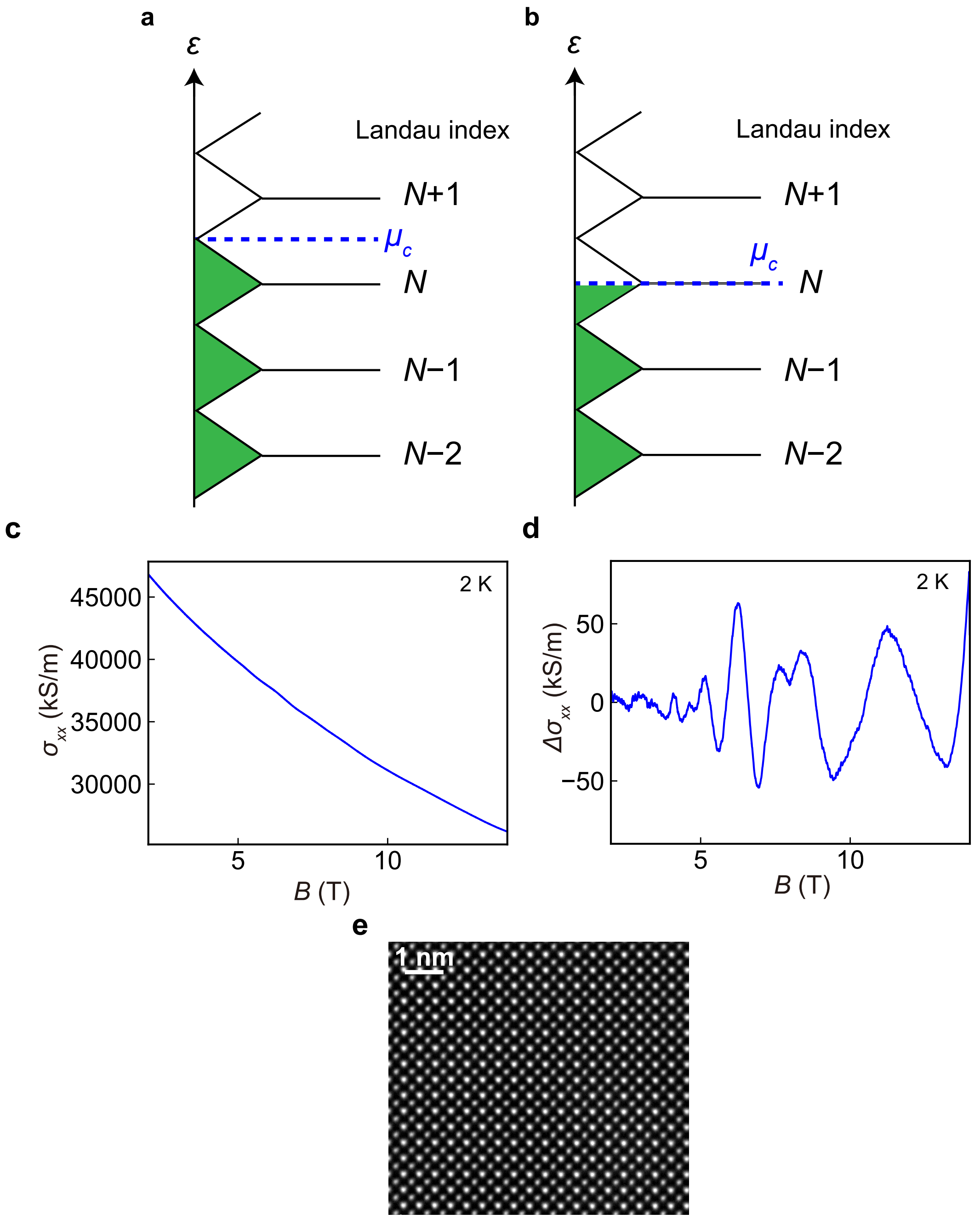

**Extended Data Fig. 5 STEM image and Landau quantization and chemical potential**

**a**, **b,** Schematic energy diagram of the relationship between Landau quantized levels $\epsilon_N$ and chemical potential $\mu_c$ which is located at the middle point between the *N*th and (*N*+1)th Landau level, and at the center of *N*th Landau level, respectively. The green region represents the filling ratio of each Landau level. In **a** and **b**, the *N*th Landau level is fully and half occupied, respectively. **c,** Raw $\sigma_{xx}(B)$ data measured at 2 K and $\beta = \gamma =$ 90° for the $SrRuO_3$ film with the RRR of 84.3. The raw conductivity data is obtained by using $\rho_{xx}(B)$ and $\rho_{xy}(B)$ data as $\sigma_{xx}(B) = \rho_{xx}(B)/((\rho_{xx}(B))^2 + (\rho_{xy}(B))^2)$. **d,** SdH oscillation data $\Delta\sigma_{xx}(B)$ obtained by subtracting a polynomial function up to the fifth order from $\sigma_{xx}(B)$ in **c**. **e,** Cross-sectional high-angle annular dark field scanning transmission electron microscopy (HAADF-STEM) image of the $SrRuO_3$ film with the RRR of 71 taken along the [100] axis of the $SrTiO_3$ substrate.

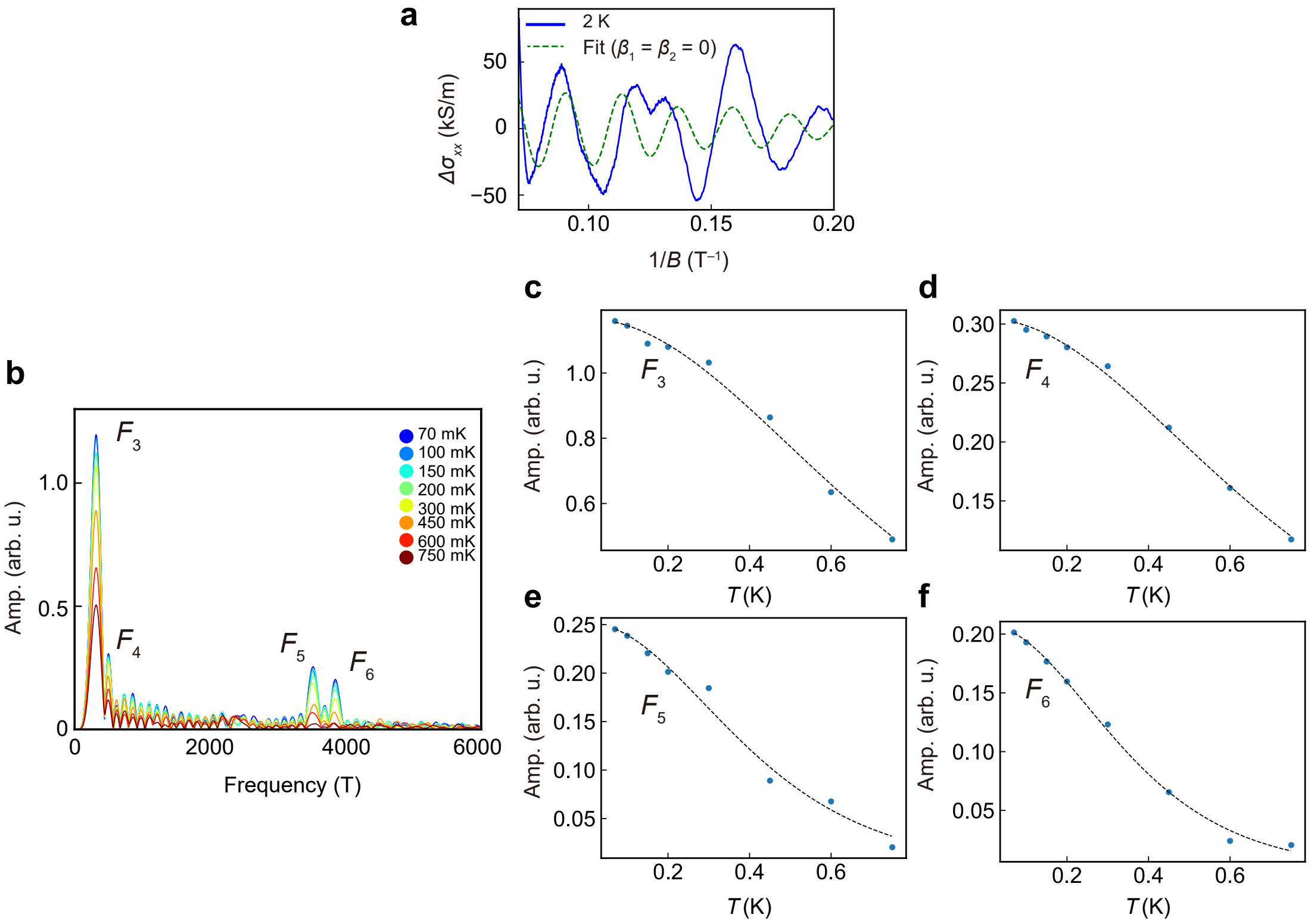

**Extended Data Fig. 6 LK theory fitting to the SdH oscillations with the fixed zero Berry phase and mass estimations of the trivial orbits. a,** SdH oscillation data at 2 K for the $SrRuO_3$ film with the RRR of 84.3 and the fitting results by eq. (1) with the zero Berry phases ($\beta_1 = \beta_2 = 0$). The SdH oscillation is the same data as in Fig. 3b. The fitting was carried out by a non-linear least squares method with the fitting parameters $A_1$, $T_{D1}$, $A_2$, and $T_{D2}$. The fitting curve cannot reproduce the experimental data well, confirming the existence of the non-zero Berry phase. **b,** Fourier transform spectra of the SdH oscillations from 70 to 750 mK for the $SrRuO_3$ film with the RRR of 84.3. The spectra are obtained by fast Fourier transform for the oscillation data $\Delta\rho_{xx}(B)$ from 12.5 T to 14 T. $F_3$, $F_4$, $F_5$, and $F_6$ peaks correspond to 300, 500, 3500, and 3850 T, respectively. **c-f,** Mass estimations for the $F_3$, $F_4$, $F_5$, and $F_6$ orbits according to the LK theory. Black dashed curves are fitting curves.

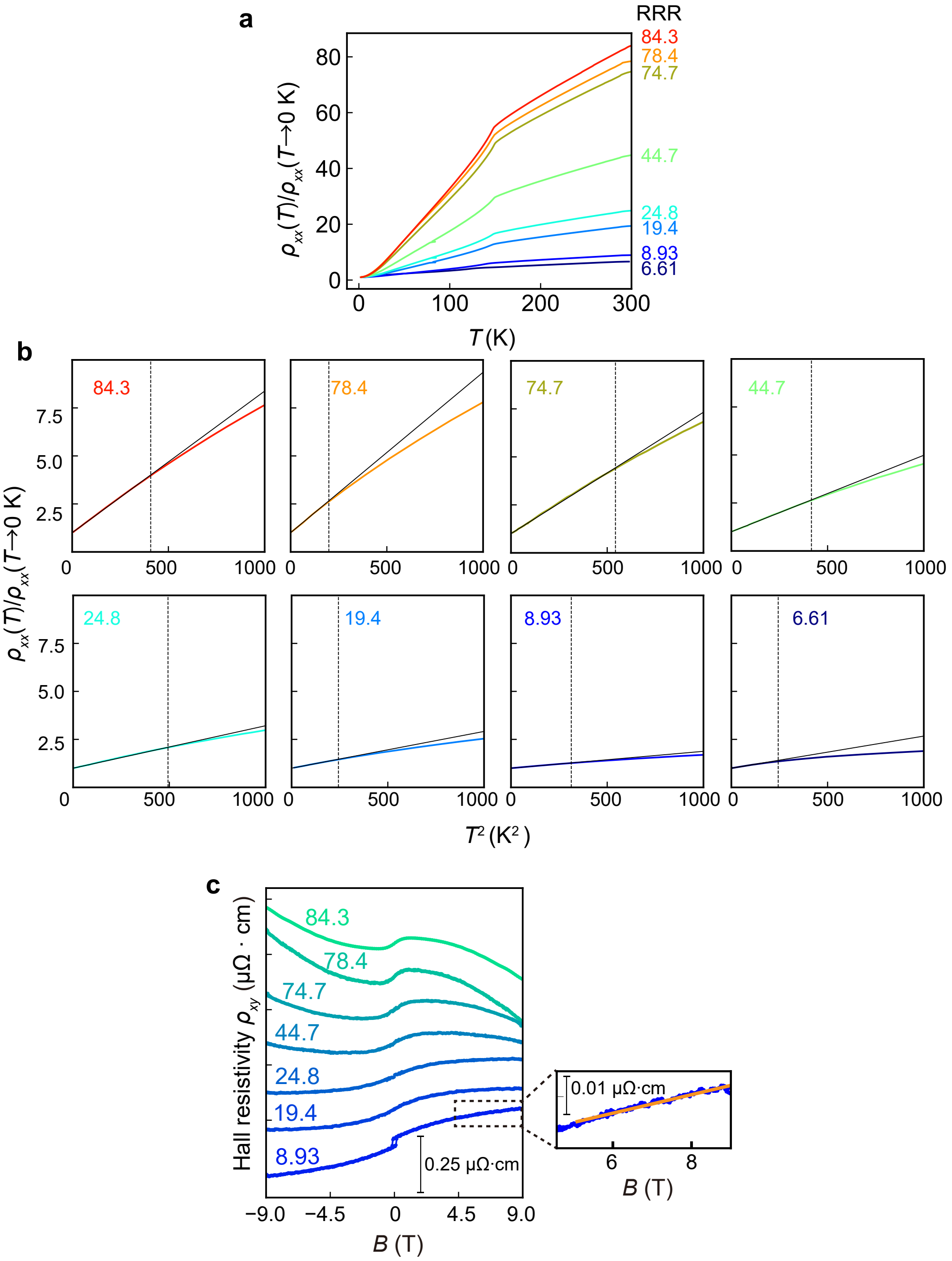

**Extended Data Fig. 7 RRR dependence of $T_C$, $T_F$, and Hall resistivity. a,** $\rho_{xx}(T)/\rho_{xx}(T\to 0$ K) versus $T$ of the different RRR samples. The kinks around 150 K correspond to $T_C$ of the sample. **b**, $\rho_{xx}(T)/\rho_{xx}$ ($T\to 0$ K) versus $T^2$ curves with the linear fittings (black lines) for the different RRR samples. The black dashed lines correspond to $T_F^2$ where the experimental $\rho_{xx}$ and the fitting line are close enough to each other (< 0.1 μΩ cm). **c,** Hall resistivity $\rho_{xy}(B)$ curves of the different RRR samples at 2 or 2.3 K with $B$ applied in the out-of-plane [001] direction of the $SrTiO_3$ substrate. The right-side figure is enlarged graph of $\rho_{xy}(B)$ with RRR = 8.93 at 4.5 T < $B$ < 9 T. The orange line is the linear fitting result from 5T to 9 T for estimating the carrier density and mobility.

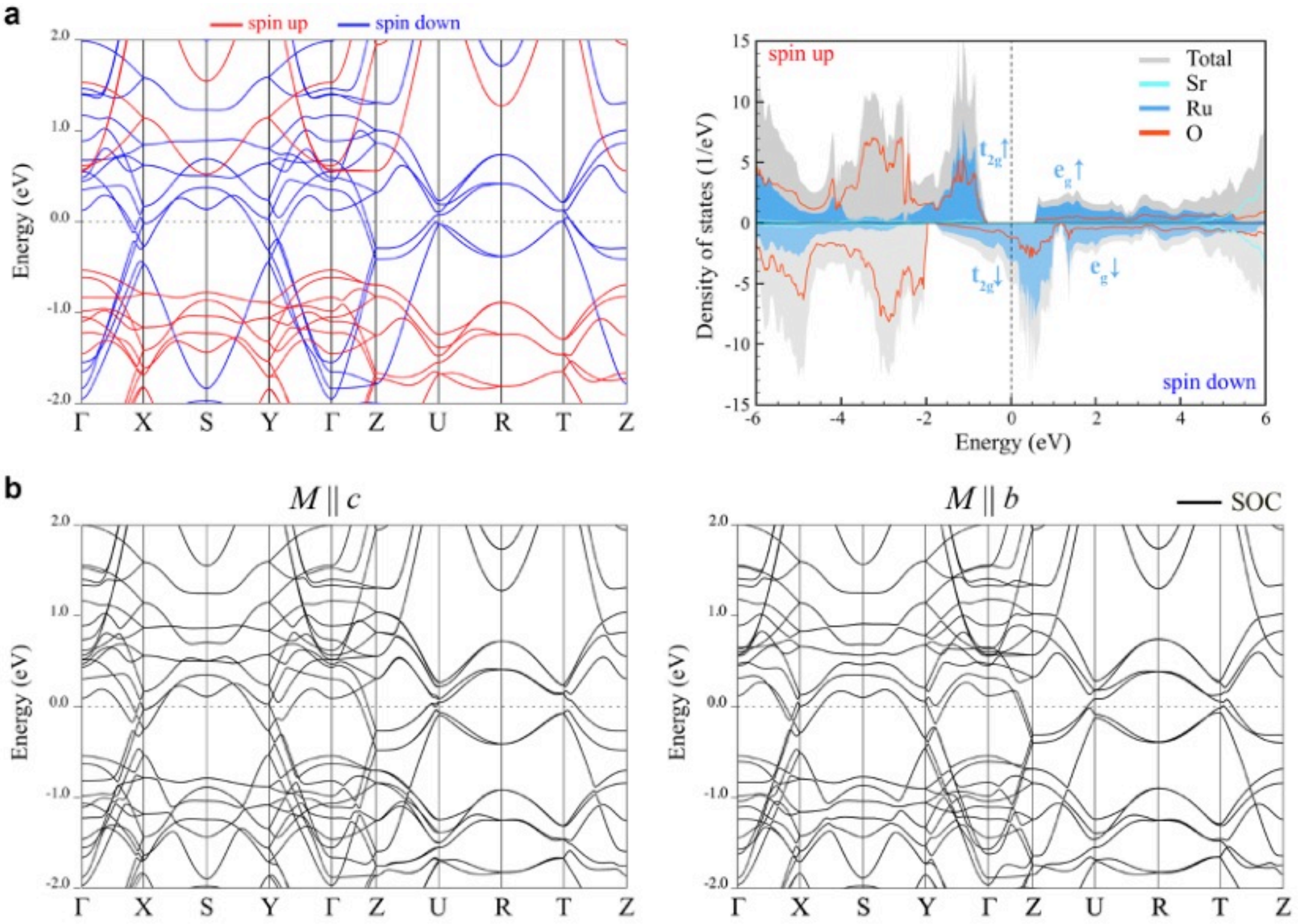

**Extended Data Fig. 8 Electronic structure of orthorhombic $SrRuO_3$ calculated within GGA+*U* with *U* = 2.6 eV and *J* = 0.6 eV. a,** Band structure (left) and density of states (right) for the ferromagnetic state without SOC. **b**, Band structure for the ferromagnetic state with SOC and the magnetization along the orthorhombic *c* and *b* axes (left and right, respectively) calculated within GGA+*U*+SOC. Fractional coordinates of the high-symmetry *k*-points are Γ (0,0,0), X (0.5,0,0), S (0.5,0.5,0), Y (0,0.5,0), Z (0,0,0.5), U (0.5,0,0.5), R (0.5,0.5,0.5), and T (0,0.5,0.5).

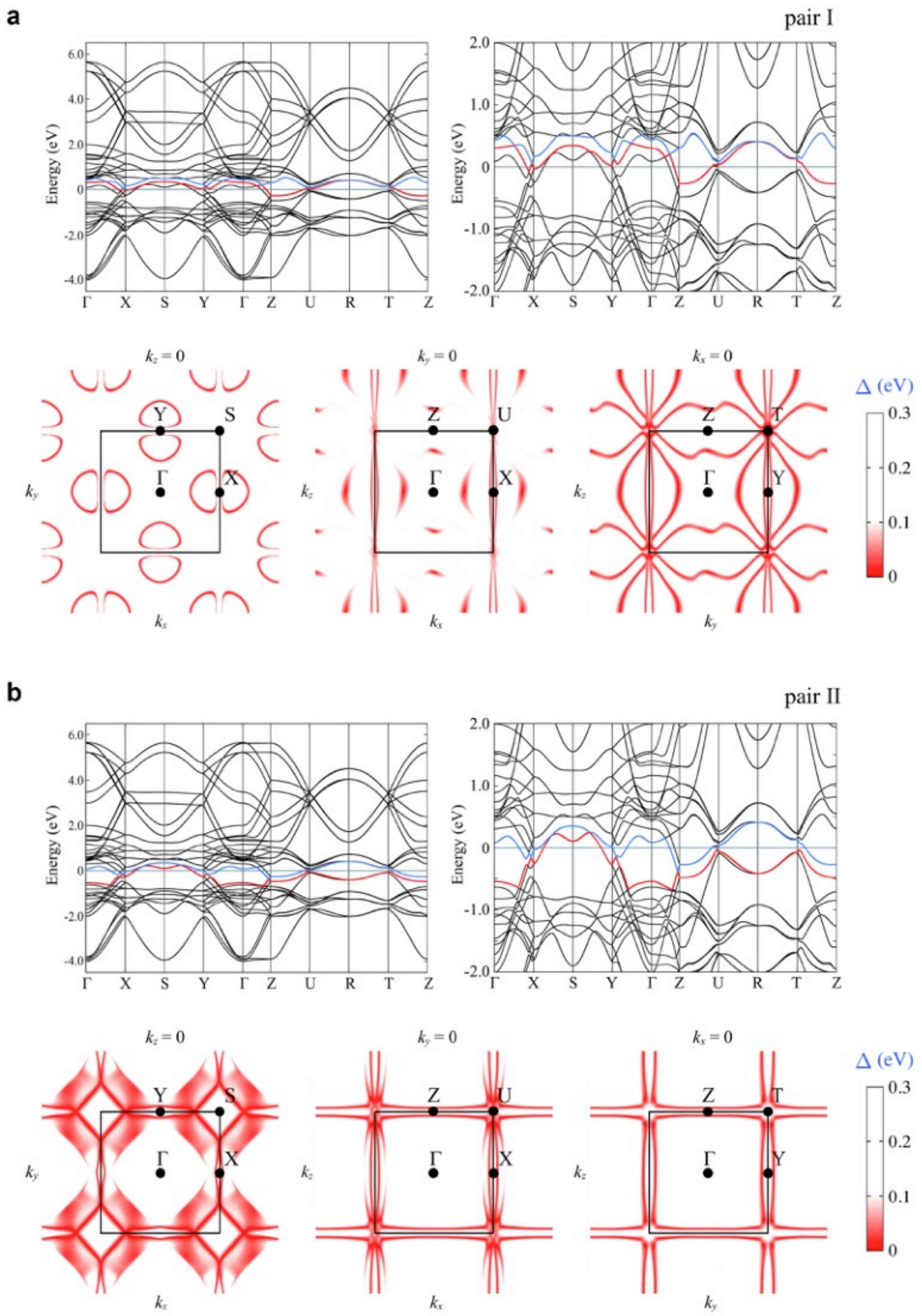

**Extended Data Fig. 9 Pairs of bands I and II in the case of magnetization along the orthorhombic *c* axis. a,** Gap function $\Delta \leq 0.1$ eV for the pair I (red and blue bands in upper figures in **a**) in the $k_z = 0$, $k_y = 0$, $k_x = 0$ planes. **b**, Gap function $\Delta \leq 0.1$ eV for the pair II (red and blue bands in upper figures in **b**) in the $k_z$=0, $k_y$=0, $k_x$=0 planes. The band structure is interpolated in the Wannier basis representing the $t_{2g}$ and $e_g$ states based on the electronic structure calculated within GGA+$U$+SOC.

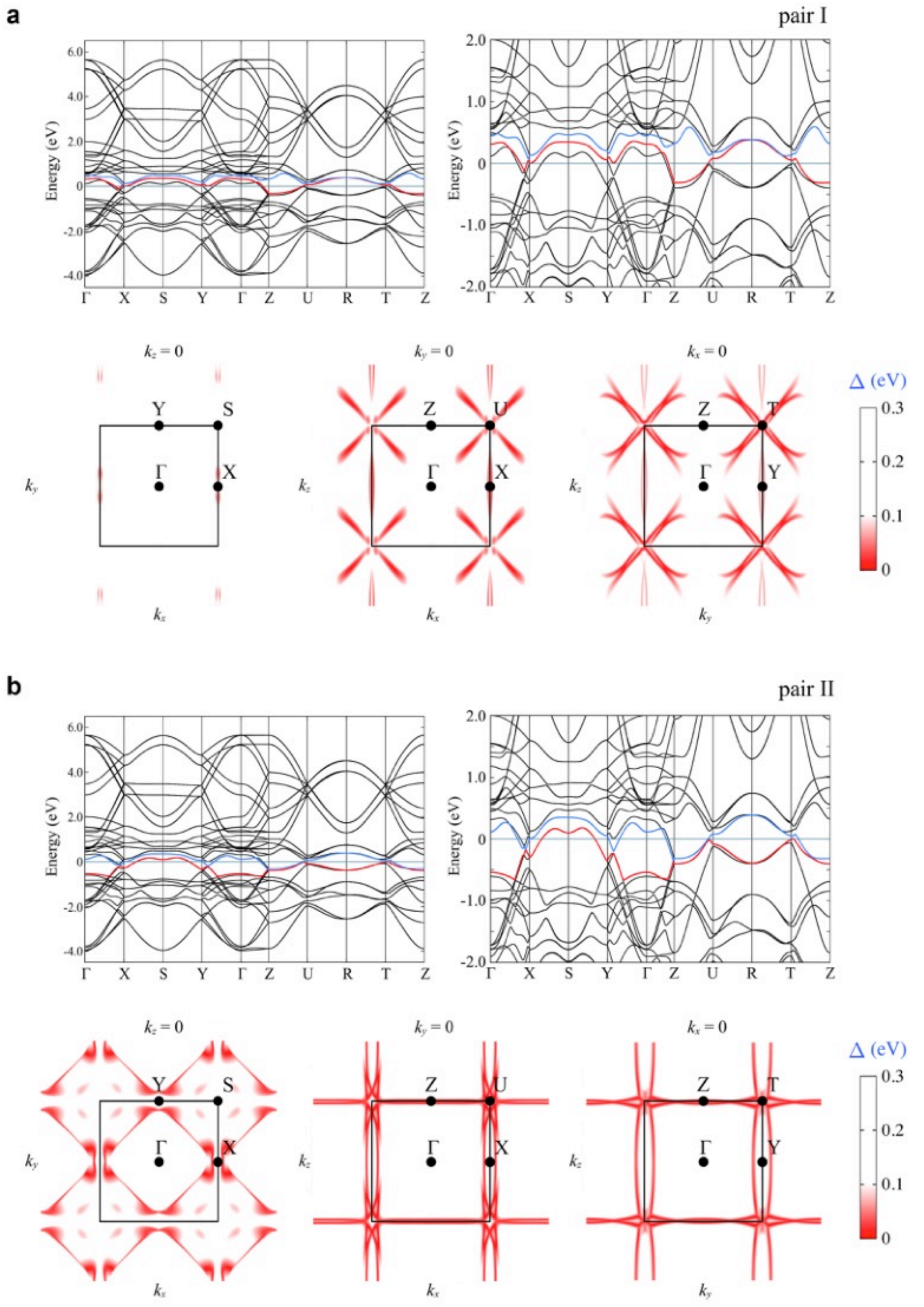

**Extended Data Fig. 10 Pairs of bands I and II in the case of magnetization along the orthorhombic *b* axis. a,** Gap function $\Delta \leq 0.1$ eV for the pair I (red and blue bands in upper figures in **a**) in the $k_z = 0$, $k_y = 0$, $k_x = 0$ planes. **b**, Gap function $\Delta \leq 0.1$ eV for the pair II (red and blue bands in upper graphs in **b**) in the $k_z = 0$, $k_y = 0$, $k_x = 0$ planes. The band structure is interpolated in the Wannier basis representing the $t_{2g}$ and $e_g$ states based on the electronic structure calculated within GGA+*U*+SOC.

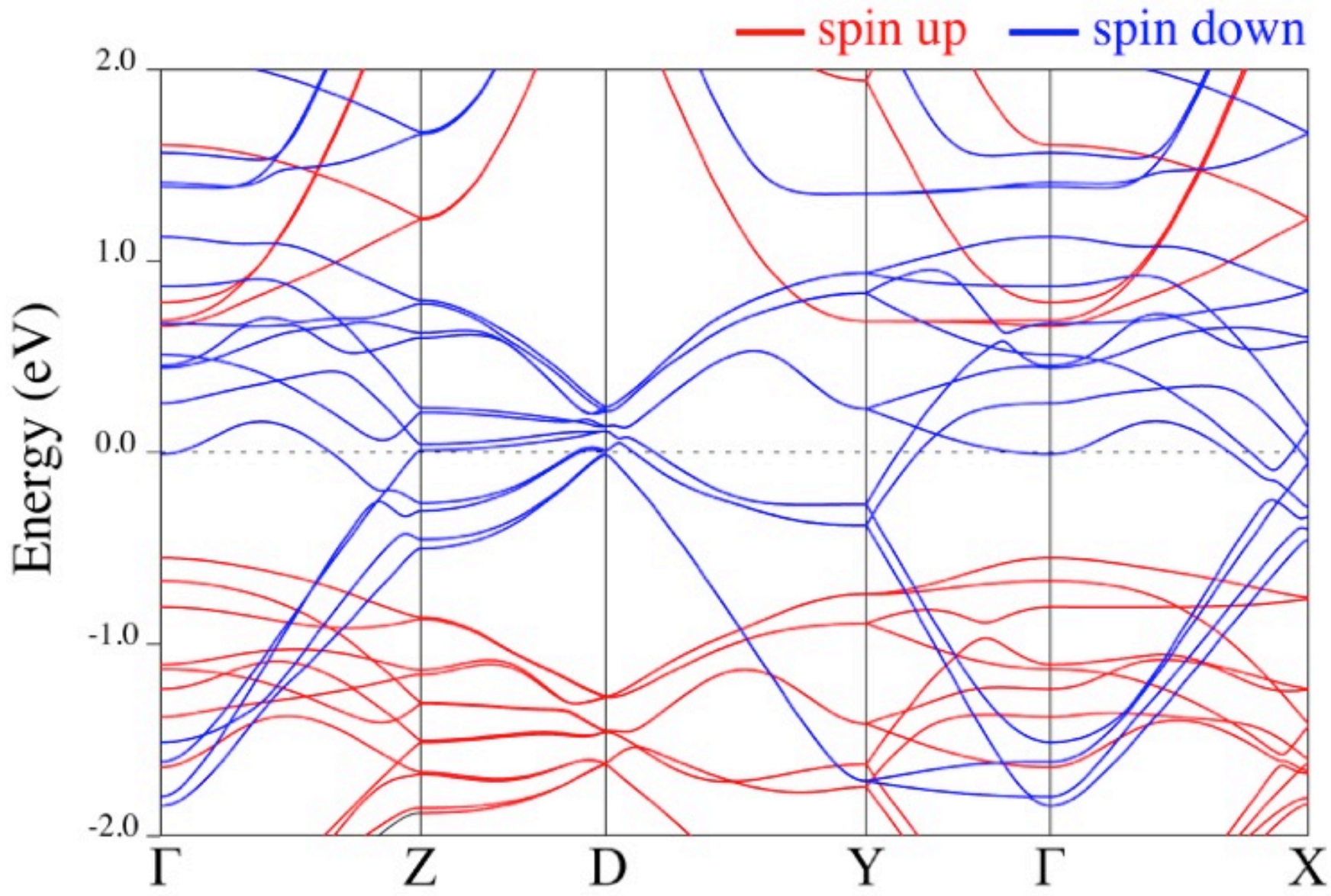

**Extended Data Fig. 11 Electronic structure of monoclinic $SrRuO_3$ in the ferromagnetic state without SOC calculated within GGA+*U* with *U* = 2.6 eV and *J* = 0.6 eV.** Fractional coordinates of the high-symmetry *k*-points are Γ (0,0,0), Z (0.5,0,0), D (0.5,0,0.5), Y (0,0,0.5), and X (0,0.5,0).

**Extended Data Table 1 Frequencies and effective cyclotron masses estimated from SdH oscillations in the $SrRuO_3$ film with RRR = 84.3.** Frequencies and effective cyclotron masses estimated in Fig. 3c and Extended Data Fig. 6c-f. $m_0$ represents the free electron mass in a vacuum.

| | $F_1$ | $F_2$ | $F_3$ | $F_4$ | $F_5$ | $F_6$ |
|---|---|---|---|---|---|---|
| $F$ (T) | 26 | 44 | 300 | 500 | 3500 | 3850 |
| $m^*/m_0$ | 0.35 | 0.58 | 2.9 | 3.1 | 5.0 | 5.8 |

**Extended Data Table 2 Magnetic moments (in $\mu_B$) of $SrRuO_3$ in the ferromagnetic state as obtained from GGA+*U*+SOC.** Second and third columns show the calculated magnetic moments in $\mu_B$ units when the magnetization is along the orthorhombic *c* and *b* axes, respectively. ***S*** and ***L*** are spin and orbital magnetic moments, respectively. Ru sites are given in fractional coordinates.

| Ru site | $M \parallel c$ | $M \parallel b$ |
|---|---|---|
| (0.5, 0, 0) | ***S*** = (0.064, 0.010, 1.399)<br>***L*** = (0, -0.005, 0.074) | ***S*** = (0.057, 1.398, -0.020)<br>***L*** = (0.003, 0.099, -0.008) |
| (0, 0.5, 0) | ***S*** = (-0.064, 0.010, 1.400)<br>***L*** = (0, -0.005, 0.074) | ***S*** = (-0.056, 1.400, -0.021)<br>***L*** = (-0.003, 0.099, -0.008) |
| (0.5, 0, 0.5) | ***S*** = (-0.064, -0.010, 1.397)<br>***L*** = (0, 0.005, 0.074) | ***S*** = (0.059, 1.397, -0.020)<br>***L*** = (0.003, 0.099, 0.008) |
| (0.5, 0, 0.5) | ***S*** = (0.064, -0.010, 1.396)<br>***L*** = (0, 0.005, 0.074) | ***S*** = (-0.057, 1.399, -0.021)<br>***L*** = (-0.003, 0.099, 0.008) |

**Extended Data Table 3 Weyl points calculated for the pair of bands I for the case of the magnetization along the *c* axis.** The *k*-points are given in fractional coordinates.

| # | $E$-$E_F$, (eV) | ($k_x$, $k_y$, $k_z$) | Chirality |
|---|---|---|---|
| $WP_z1_1$ | 0.204 | (0, 0, 0.332) | 1 |
| $WP_z1_2$ | 0.204 | (0, 0, -0.332) | -1 |
| $WP_z2_1$ | 0.214 | (0, 0.307, 0.357) | -1 |
| $WP_z2_2$ | 0.219 | (0, 0.305, -0.355) | 1 |
| $WP_z2_3$ | 0.219 | (0, -0.305, 0.355) | -1 |
| $WP_z2_4$ | 0.214 | (0, -0.307, -0.357) | 1 |
| $WP_z3_1$ | 0.117 | (0, 0.485, 0.456) | -1 |
| $WP_z3_2$ | 0.116 | (0, -0.484, 0.456) | -1 |
| $WP_z3_3$ | 0.116 | (0, 0.484, -0.456) | 1 |
| $WP_z3_4$ | 0.117 | (0, -0.485, -0.456) | 1 |
| $WP_z4_1$ | 0.102 | (-0.022, 0.439, -0.478) | -1 |
| $WP_z4_2$ | 0.102 | (0.020, 0.439, -0.478) | -1 |
| $WP_z4_3$ | 0.101 | (-0.027, 0.437, 0.479) | 1 |
| $WP_z4_4$ | 0.101 | (0.029, 0.436, 0.479) | 1 |
| $WP_z4_5$ | 0.102 | (-0.021, -0.439, 0.479) | 1 |
| $WP_z4_6$ | 0.102 | (0.023, -0.438, 0.479) | 1 |
| $WP_z4_7$ | 0.101 | (0.027, -0.437, -0.479) | -1 |
| $WP_z4_8$ | 0.101 | (-0.029, -0.436, -0.479) | -1 |
| $WP_z5_1$ | 0.089 | (0.235, -0.306, 0.483) | -1 |
| $WP_z5_2$ | 0.087 | (0.233, 0.306, 0.483) | -1 |
| $WP_z5_3$ | 0.087 | (0.233, -0.306, -0.483) | 1 |
| $WP_z5_4$ | 0.089 | (0.235, 0.306, -0.483) | 1 |
| $WP_z5_5$ | 0.087 | (-0.233, 0.306, 0.483) | -1 |
| $WP_z5_6$ | 0.087 | (-0.233, -0.306, -0.483) | 1 |
| $WP_z5_7$ | 0.089 | (-0.235, 0.306, -0.483) | 1 |
| $WP_z5_8$ | 0.089 | (-0.235, -0.306, 0.483) | -1 |

**Extended Data Table 4 Weyl points calculated for the pair of bands II for the case of the magnetization along the *c* axis.** The *k*-points are given in fractional coordinates.

| # | $E$-$E_F$, (eV) | ($k_x$, $k_y$, $k_z$) | Chirality |
|---|---|---|---|
| $WP_z6_1$ | -0.008 | (0, 0.464, 0.400) | 1 |
| $WP_z6_2$ | -0.016 | (0, 0.464, -0.400) | -1 |
| $WP_z6_3$ | -0.016 | (0, -0.464, 0.400) | 1 |
| $WP_z6_4$ | -0.008 | (0, -0.464, -0.400) | -1 |
| $WP_z7_1$ | -0.104 | (0.452, 0, 0.385) | -1 |
| $WP_z7_2$ | -0.104 | (-0.452, 0, -0.385) | 1 |
| $WP_z7_3$ | -0.122 | (0.450, 0, -0.368) | 1 |
| $WP_z7_4$ | -0.122 | (-0.450, 0, 0.368) | -1 |
| $WP_z8_1$ | 0.239 | (0.276, 0.485, -0.205) | 1 |
| $WP_z8_2$ | 0.225 | (0.266, 0.483, 0.223) | -1 |
| $WP_z8_3$ | 0.220 | (0.262, -0.483, 0.228) | -1 |
| $WP_z8_4$ | 0.231 | (0.270, -0.484, -0.215) | 1 |
| $WP_z8_5$ | 0.238 | (-0.275, -0.485, 0.209) | -1 |
| $WP_z8_6$ | 0.230 | (-0.269, 0.484, 0.218) | -1 |
| $WP_z8_7$ | 0.220 | (-0.262, 0.483, -0.228) | 1 |
| $WP_z8_8$ | 0.227 | (-0.267, -0.484, -0.220) | 1 |
| $WP_z9_1$ | 0.142 | (-0.492, 0.187, 0.170) | -1 |
| $WP_z9_2$ | 0.142 | (0.492, 0.188, 0.170) | -1 |
| $WP_z9_3$ | 0.151 | (0.492, -0.191, 0.175) | -1 |
| $WP_z9_4$ | 0.149 | (-0.492, -0.191, 0.174) | -1 |
| $WP_z9_5$ | 0.142 | (-0.491, 0.193, -0.176) | 1 |
| $WP_z9_6$ | 0.142 | (-0.492, -0.188, -0.171) | 1 |
| $WP_z9_7$ | 0.149 | (0.492, -0.188, -0.171) | 1 |
| $WP_z9_8$ | 0.150 | (0.492, 0.192, -0.175) | 1 |
| $WP_z10_1$ | 0.257 | (0.408, -0.379, -0.121) | -1 |
| $WP_z10_2$ | 0.257 | (0.426, 0.362, -0.123) | -1 |
| $WP_z10_3$ | 0.257 | (0.425, -0.363, 0.124) | 1 |
| $WP_z10_4$ | 0.258 | (-0.412, 0.374, 0.128) | 1 |
| $WP_z10_5$ | 0.258 | (-0.412, -0.375, -0.129) | -1 |
| $WP_z10_6$ | 0.256 | (-0.427, 0.360, -0.116) | -1 |
| $WP_z10_7$ | 0.258 | (0.403, 0.384, 0.119) | 1 |
| $WP_z10_8$ | 0.256 | (-0.428, -0.359, 0.116) | 1 |

**Extended Data Table 5 Weyl points calculated for the pair of bands I for the case of the magnetization along the *b* axis.** The *k*-points are given in fractional coordinates.

| # | $E$-$E_F$, (eV) | ($k_x$, $k_y$, $k_z$) | Chirality |
|---|---|---|---|
| $WP_y1_1$ | 0.160 | (0, 0.378, 0.395) | 1 |
| $WP_y1_2$ | 0.152 | (0, 0.379, -0.394) | 1 |
| $WP_y1_3$ | 0.160 | (0, -0.378, -0.395) | -1 |
| $WP_y1_4$ | 0.152 | (0, -0.379, 0.394) | -1 |
| $WP_y2_1$ | 0.126 | (0.485, -0.095, 0) | 1 |
| $WP_y2_2$ | 0.128 | (-0.486, -0.097, 0) | 1 |
| $WP_y2_3$ | 0.128 | (0.486, 0.097, 0) | -1 |
| $WP_y2_4$ | 0.152 | (-0.485, 0.095, 0) | -1 |
| $WP_y3_1$ | 0.275 | (0, -0.326, -0.297) | -1 |
| $WP_y3_2$ | 0.275 | (0, 0.326, 0.297) | 1 |
| $WP_y3_3$ | 0.282 | (0, 0.321, -0.294) | 1 |
| $WP_y3_4$ | 0.281 | (0, -0.321, 0.294) | -1 |
| $WP_y4_1$ | 0.168 | (0.323, -0.081, -0.375) | -1 |
| $WP_y4_2$ | 0.163 | (0.331, 0.076, -0.379) | 1 |
| $WP_y4_3$ | 0.165 | (0.329, -0.077, 0.377) | -1 |
| $WP_y4_4$ | 0.168 | (-0.323, 0.082, 0.375) | 1 |
| $WP_y4_5$ | 0.165 | (-0.329, 0.077, -0.377) | 1 |
| $WP_y4_6$ | 0.163 | (-0.331, -0.076, 0.379) | -1 |
| $WP_y4_7$ | 0.177 | (0.306, 0.092, 0.368) | 1 |
| $WP_y4_8$ | 0.177 | ( -0.306, -0.092, -0.368) | -1 |
| $WP_y5_1$ | 0.113 | (-0.172, 0.308, -0.437) | -1 |
| $WP_y5_2$ | 0.112 | (0.174, -0.306, 0.437) | 1 |
| $WP_y5_3$ | 0.098 | (0.196, -0.307, -0.452) | 1 |
| $WP_y5_4$ | 0.103 | (0.191, 0.305, 0.449) | -1 |
| $WP_y5_5$ | 0.099 | (-0.196, 0.303, 0.451) | -1 |
| $WP_y5_6$ | 0.102 | (-0.193, -0.304, -0.450) | 1 |
| $WP_y5_7$ | 0.096 | (-0.201, -0.302, 0.454) | 1 |
| $WP_y5_8$ | 0.096 | (0.201, 0.302, -0.454) | -1 |

**Extended Data Table 6 Weyl points calculated for the pair of bands II for the case of the magnetization along the *b* axis.** The *k*-points are given in fractional coordinates.

| # | $E$-$E_F$, (eV) | ($k_x$, $k_y$, $k_z$) | Chirality |
|---|---|---|---|
| $WP_y6_1$ | -0.070 | (0, 0.441, -0.349) | 1 |
| $WP_y6_2$ | -0.060 | (0, 0.441, 0.350) | 1 |
| $WP_y6_3$ | -0.070 | (0, -0.441, 0.349) | -1 |
| $WP_y6_4$ | -0.060 | (0, -0.441, -0.349) | -1 |
| $WP_y7_1$ | -0.318 | (0.155, -0.328, 0) | 1 |
| $WP_y7_2$ | -0.331 | (0.169, 0.313, 0) | -1 |
| $WP_y7_3$ | -0.320 | (-0.157, 0.326, 0) | -1 |
| $WP_y7_4$ | -0.331 | (-0.168, -0.314, 0) | 1 |
| $WP_y8_1$ | -0.243 | (-0.398, -0.101, 0) | -1 |
| $WP_y8_2$ | -0.260 | (-0.391, 0.104, 0) | 1 |
| $WP_y8_3$ | -0.240 | (0.399, 0.100, 0) | 1 |
| $WP_y8_4$ | -0.260 | (0.390, -0.104, 0) | -1 |
| $WP_y9_1$ | -0.160 | (0.454, 0.020, 0) | -1 |
| $WP_y9_2$ | -0.160 | (-0.454, -0.021, 0) | 1 |
| $WP_y9_3$ | -0.153 | (-0.455, 0.026, 0) | -1 |
| $WP_y9_4$ | -0.154 | (0.455, -0.025, 0) | 1 |

**References**

60. Wakabayashi, Y. K. *et al*. Ferromagnetism above 1000 K in a highly cation-ordered double-perovskite insulator $Sr_3OsO_6$. *Nat. Commun.* **7**, 101114 (2019).
61. Snoek, J., Larochelle, H. & Adams, R. P. Practical Bayesian Optimization of Machine Learning Algorithms. *Advances in Neural Information Processing Systems* 25 (2012).
62. Wakabayashi, Y. K. *et al*. Improved adaptive sampling method utilizing Gaussian process regression for prediction of spectral peak structures. *Appl. Phys. Express* **11**, 112401 (2018).
63. Mockus, J. Tiesis, V. & Zilinskas, A. The application of Bayesian methods for seeking the extremum. *Towards Global Optimization* **2**, 117 (1978).
64. Collaudin, A. *et al*. Angle dependence of the orbital magnetoresistance in bismuth. *Phys. Rev. X* **5**, 021022 (2015).
65. Kim, H. *et al*, Dirac versus weyl fermions in topological insulators: Adler-Bell-Jackiw anomaly in transport phenomena. *Phys. Rev. Lett.* **111**, 246603 (2013).
66. Rötger, T. *et al*. Relation between low-temperature quantum and high-temperature classical magnetotransport in a two-dimensional electron gas. *Phys. Rev. Lett.* **62**, 90 (1989).
67. Parish, M. M. & Littlewood, P. B. Non-saturating magnetoresistance in heavily disordered semiconductors. *Nature* **426**, 162 (2003)
68. Wang, W. J. *et al*. Disorder-dominated linear magnetoresistance in topological insulator $Bi_2Se_3$ thin films. *Appl. Phys. Lett.* **111**, 232105 (2017)
69. Abrikosov, A. Quantum magnetoresistance. *Phys. Rev. B* **58**, 2788 (1998).
70. Abrikosov, A. Quantum linear magnetoresistance. *Euro. Phys. Lett.* **49**, 789 (2000).
71. Hu, J. & Rosenbaum, T. Classical and quantum routes to linear magnetoresistance. *Nat. Mater.* **7**, 697 (2008).
72. Murayama, Y. *Mesoscopic Systems: Fundamentals and Applications* (Wiley-VCH,Weinheim, Germany, 2001), pp. 213.
73. Nagaosa, N. *et al.* Anomalous Hall effect. *Rev. of Mod. Phys.* **82**, 1539 (2010).
74. Raquet, B. *et al.* Electron-magnon scattering and magnetic resistivity in 3d ferromagnets. *Phys. Rev. B* **66**, 024433 (2002).
75. Jenni, K. *et al*., Interplay of Electronic and Spin Degrees in Ferromagnetic $SrRuO_3$: Anomalous Softening of the Magnon Gap and Stiffness. *Phys. Rev. Lett.* **123**, 017202 (2019).
76. Nikolaeva, A. *et al*. Diameter-dependent thermopower of bismuth nanowires. *Phys. Rev. B* **77**, 035422 (2008).
77. Kuntsevich A. Yu., Shupletsov, A. V., and Minkov, G. M. Simple mechanisms that impede the Berry phase identification from magneto-oscillations. *Phys. Rev. B* **97**, 195431 (2018).
78. Beukman, A. *et al*. Spin-orbit interaction in a dual gated InAs/GaSb quantum well. *Phys. Rev. B* **96** 241401(R) (2017).
79. Nichele, F. *Transport experiments in two-dimensional systems with strong spin-orbit interaction.* PhD thesis, Eidgenössische Technische Hochschule Zürich, (2014).
80. Shoenberg, D. Magnetic interaction and phase smearing. *J. Low Temp. Phys.* **25**, 755 (1976).

81. Itskovsky, M. A. Kventsel, & G. F. Maniv, T. Periodic diamagnetic domain structures in metals under a quantizing magnetic field. *Phys. Rev. B* **50**, 6779 (1994).
82. Pippard, A. B. Quantization of coupled orbits in metals. *Proc. R. Soc. London, Ser. A* **270**, 1340 (1963)
83. Shoenberg, D. *Magnetic Oscillations in Metals* (Cambridge University Press, Cambridge, UK, 1984).
84. Masuda, H. *et al.* Impact of antiferromagnetic order on Landau-level splitting of quasi-two-dimensional Dirac fermions in $EuMnBi_2$. *Phys. Rev. B* **98**, 161108(R) (2018)
85. Hu, J. *et al.* π Berry phase and Zeeman splitting of Weyl semimetal TaP. *Sci. Rep.* **6**, 18674 (2016)
86. Ryee, S. *et al.* Quasiparticle self-consistent GW calculation of $Sr_2RuO_4$ and $SrRuO_3$. *Phys. Rev. B* **93**, 075125 (2016).
87. Lifshitz, I. M. & Kosevich, A. M. Theory of Magnetic Susceptibility in Metals at Low Temperatures. *Sov. Phys. JETP* **2**, 636 (1956).
88. Katano, T & Tsuda, N. The Seebeck Coefficient of $Ca_{1-x}Sr_xRuO_3$. *J. Phys. Soc. Jpn.* **65**, 207 (1996).
89. Perdew, J. P., Burke, K., & Ernzerhof, M. Generalized Gradient Approximation Made Simple. *Phys. Rev. Lett.* **77**, 3865 (1996).
90. Kresse, G. & Joubert, D. From ultrasoft pseudopotentials to the projector augmented-wave method. *Phys. Rev. B* **59**, 1758 (1999).
91. Kresse, G. & Furthmüller, J. Efficient iterative schemes for ab initio total-energy calculations using a plane-wave basis set. *Phys. Rev. B* **54**, 11169 (1996).
92. Monkhorst, H. J. & Pack, James, D. Special points for Brillouin-zone integrations. *Phys. Rev. B* **13**, 5188 (1976).
93. Liechtenstein, A.I., Anisimov, V.I., & Zaanen, J. Density-functional theory and strong interactions: Orbital ordering in Mott-Hubbard insulators. *Phys. Rev. B* **52**, R5467 (1995).
94. Solovyev, I.V., Dederichs, P. H., & Anisimov, V. I. Corrected atomic limit in the local-density approximation and the electronic structure of d impurities in Rb. *Phys. Rev. B* **50**, 16861 (1994).
95. Rondinelli, J. M., Caffrey, N. M., Sanvito, S., & Spaldin, N. A. Electronic properties of bulk and thin film $SrRuO_3$: Search for the metal-insulator transition. *Phys. Rev. B* **78**, 155107 (2008).
96. Grånäs, O., *et al.* Electronic structure, cohesive properties, and magnetism of $SrRuO_3$. *Phys. Rev. B* **90**, 165130 (2014).
97. Mostofi, A. A., *et al.* An updated version of wannier90: A tool for obtaining maximally-localised Wannier functions. *Comput. Phys. Commun* **185**, 2309 (2014).
98. Gosálbez-M, D. Souza, I., & Vanderbilt, D. Chiral degeneracies and Fermi-surface Chern numbers in bcc Fe. *Phys. Rev. B* **92**, 085138 (2015).